\documentclass[aps,prx,twocolumn,superscriptaddress,showpacs,nofootinbib]{revtex4-2}
\usepackage{amsmath}
\usepackage{amssymb,amsfonts}
\usepackage{amsthm,amsmath,amssymb}
\usepackage{dsfont}
\usepackage{mathrsfs}

\usepackage{graphicx}
\usepackage[colorlinks]{hyperref}
\usepackage{dsfont,bbm}
\usepackage{amsfonts,amssymb}

\usepackage{subfigure}

\usepackage{multirow}
\usepackage{youngtab}

\usepackage{caption}

\usepackage{extarrows}

\usepackage[textsize=footnotesize,textwidth=2.25cm]{todonotes}

\newcommand\symbolOa{\mathcal{O}}
\newcommand\symbolOb{\mathcal{O}}

\newcommand\smallN{\sigma}

\newcommand{\operD}{\mathbb{D}}
\newcommand{\operH}{\mathbb{H}}

\graphicspath{{./figs/}}

\begin{document}

\begin{flushright}
  \small USTC-ICTS/PCFT-26-20
\end{flushright}
\vspace{-1.5em}  
\title{Non-Hermitian Structure and Exceptional Points in Yang-Mills Theory \\ from Analytic Continuation of $N_c$}

\affiliation{School of Physics and Astronomy, Sun Yat-Sen University, Zhuhai 519082, China}
\author{Qingjun Jin}
\affiliation{Graduate School of China Academy of Engineering Physics, \\
No.~10 Xibeiwang East Road, Haidian District, Beijing, 100193, China}
\author{Ke Ren}
\email{renk9@mail.sysu.edu.cn}
\affiliation{School of Physics and Astronomy, Sun Yat-Sen University, Zhuhai 519082, China}
\author{Gang Yang}
\email{yangg@itp.ac.cn}
\affiliation{Institute of Theoretical Physics, Chinese Academy of Sciences, Beijing 100190, China}
\affiliation{School of Fundamental Physics and Mathematical Sciences, Hangzhou Institute for Advanced Study, UCAS, Hangzhou 310024, China}
\affiliation{School of Physical Sciences, University of Chinese Academy of Sciences, Beijing 100049, China}
\affiliation{Peng Huanwu Center for Fundamental Theory, Hefei, Anhui 230026, China}
\author{Rui Yu}
\affiliation{School of Physical Science and Technology, Inner Mongolia University, Hohhot 010021, China}
\date{\today}


\begin{abstract}
We show that analytic continuation of the number of colors, $N_c$, naturally endows Yang-Mills theory with a non-Hermitian structure. By examining the spectrum of the dilatation operator as a function of complex $N_c$, we identify a network of Exceptional Points (EPs)---non-Hermitian degeneracies where anomalous dimensions degenerate and operator eigenstates coalesce. We demonstrate that these EPs act as topological defects in complex $N_c$-space, generating non-Abelian geometric phases and enforcing nontrivial monodromies among gauge-invariant operators. Moreover, we establish a correspondence between the spontaneous breaking of an emergent $\mathcal{PT}$ symmetry of the dilatation operator and the fundamental spacetime $\mathcal{PT}$ symmetry of the underlying gauge theory. In the vicinity of EPs, the resulting non-Hermitian dynamics produces logarithmic scaling behavior in correlation functions, characteristic of logarithmic conformal field theories. Our results place conventional unitary Yang–Mills theory within a broader complexified parameter space possessing rich topological structure, suggesting a new interface between non-Hermitian physics and quantum field theory.
\end{abstract}
\maketitle

\section{Introduction}

Analytic continuation has played a foundational role in modern quantum field theories (QFTs). A classic example is the Regge theory of complex angular momentum \cite{Regge:1959mz}, which played a pivotal role in the early understanding of strong interactions. Similarly, in dimensional regularization, the spacetime dimension is treated as a complex variable to regularize divergences \cite{tHooft:1972tcz}.  

The analytic continuation of discrete group parameters has led to profound insights in statistical physics. The limit $N\rightarrow 0$ of the O($N)$ vector model revealed the physics of self-avoiding polymers \cite{deGennes:1972zz}, and its extension to continuous $N$ led to the exact solution of the loop models \cite{Nienhuis:1982fx}.
In these systems, treating the symmetry group rank as a continuous variable was not merely a mathematical convenience but a necessity for connecting disparate physical phenomena. 

In the context of high-energy theory, the rank parameter of the gauge group, $N_c$, has been treated as a continuous parameter in the large-$N_c$ expansion \cite{'tHooft:1973jz}. However, the parameter space spanned by \emph{finite} complex $N_c$ remains largely unexplored. 

In this paper, we consider the analytic continuation of the rank parameter $N_c$ in $SU(N_c)$ Yang-Mills (YM) theory.
Via the study of operator inner products and their spectrum, we demonstrate that varying $N_c$ reveals rich structures driven by non-Hermitian dynamics.
A notable feature is the appearance of Exceptional Points (EPs). 

Mathematically, an exceptional point is a singularity in the parameter space of a system where two or more eigenvalues become equal \cite{kato2013perturbation}. Crucially, their corresponding eigenvectors also coalesce, rendering the operator non-diagonalizable and giving rise to a Jordan block structure. This feature distinguishes EPs sharply from the regular degeneracy of Hermitian eigenstates.

\begin{figure*}
  \centering
  \includegraphics[width=.95\linewidth]{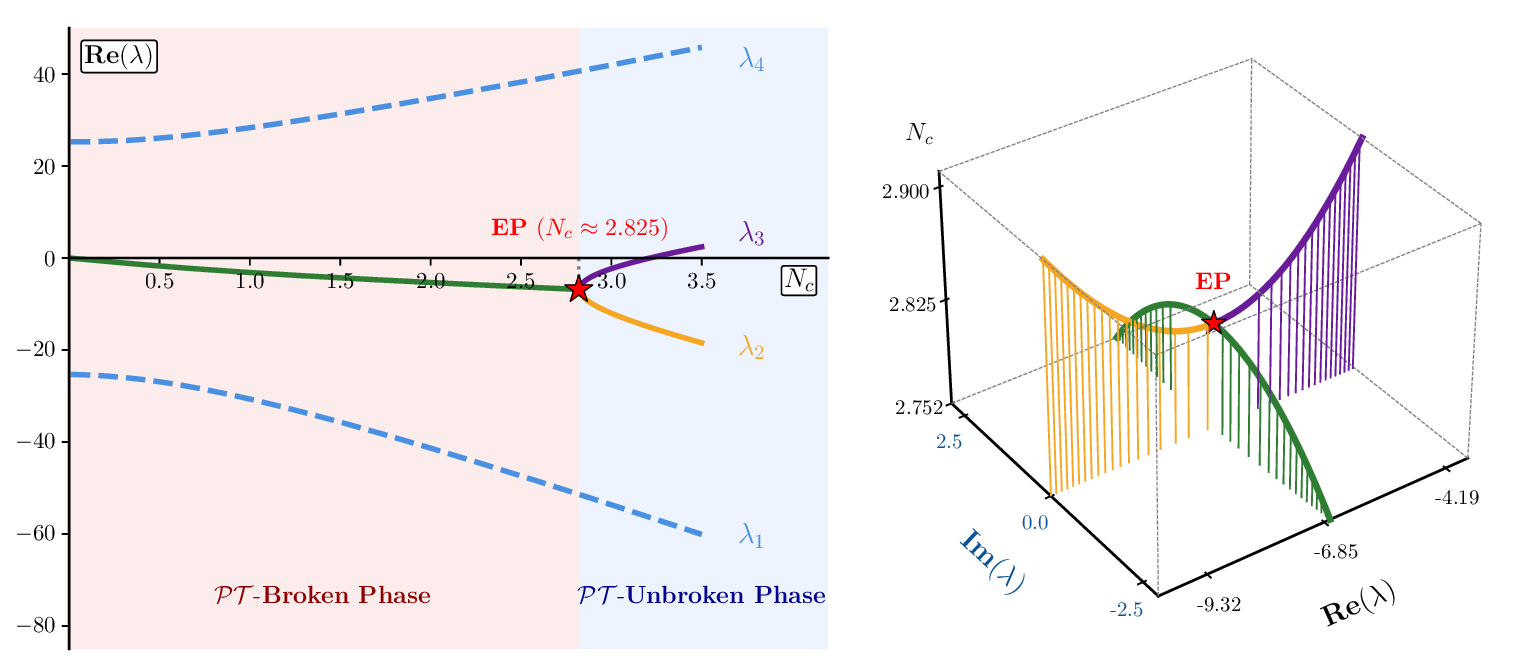}
  \caption{Operator spectrum and exceptional point for the YM operators of dimension-8 length-4 sector.} 
  \label{fig:d8L4ex}
 \end{figure*}

Exceptional points have been a central topic in non-Hermitian physics, fundamentally altering our understanding of open quantum systems \cite{Heiss:1998bv}, parity-time ($\mathcal{PT}$)-symmetric quantum mechanics \cite{Bender:1998ke, Bender:2002vv}, and pseudo-Hermitian operator structures \cite{Mostafazadeh:2001jk, Mostafazadeh:2008pw}. Recently, the study of EPs has driven a renaissance in topological physics \cite{Shen:2018cjc, Gong:2018uyu, Bergholtz:2019deh}. While most applications are in quantum-mechanical systems and optics (see e.g., the textbook \cite{Moiseyev2011} and more recent reviews in \cite{Ashida:2020dkc, Bender:2023cem}), the connection between EPs and quantum field theories is less explored. 
$\mathcal{PT}$ symmetry has been studied in specific non-Hermitian QFTs, such as the Lee model \cite{Bender:2004sv} and Fishnet theory \cite{Kazakov:2022dbd} (see also \cite{Bender:2021fxa} for the study of renormalization effects in QFTs).
The novelty of the present work is that we do not begin with a non-Hermitian theory; rather, non-Hermitian structures emerge from the analytic continuation of a unitary gauge theory. 

To study the dependence on $N_c$, we consider the set of full-color operators in SU$(N_c)$ YM theory.
The phenomenon of EPs is closely related to the existence of \emph{color-evanescent operators}.
Such operators vanish at specific integer $N_c$ due to trace identities but are non-zero at generic values.
They contribute to the unitarity violation analogous to dimensionally evanescent operators \cite{Hogervorst:2015akt, Ji:2018yaf, Jin:2023cce, Jin:2023fbz},  but are sensitive to $N_c$ rather than the spacetime dimension $d$. 

Utilizing efficient on-shell methods, we calculate the leading-order Gram matrices (associated with two-point correlation functions),
and the one- and two-loop full-color form factors.  We extract the renormalization matrices of YM operators from the form factors. 
Based on these results, we discuss how color-evanescent operators generate negative norms. 
Furthermore, complex anomalous dimensions appear or disappear when $N_c$ crosses certain critical values (typically non-integer), as shown in Figure~\ref{fig:d8L4ex}.
At these transition points, not only does a pair of anomalous dimensions degenerate, but the eigenvectors also coalesce, providing a direct signature of EPs. 
In the complex $N_c$ plane, the sheets of anomalous dimensions form Riemann surfaces with branch cuts originating from these EPs.

The emergence of EPs reveals previously unexplored structures in Yang–Mills theory.
We identify three profound physical implications of this structure. 

First, EPs mark the critical points for the spontaneous breaking of an underlying Parity-Time ($\mathcal{PT}$) symmetry, a phenomenon well-known in non-Hermitian quantum mechanics \cite{Bender:2002vv,Bender:2023cem}.
Crucially, we demonstrate that the `emergent' $\mathcal{PT}$ symmetry of the dilatation operator has a direct correspondence to the fundamental spacetime $\mathcal{PT}$-symmetry of the gauge theory.

Second, the non-Hermiticity leaves a distinct imprint on the behavior of correlation functions. In the $\mathcal{PT}$-broken phase, the complex anomalous dimensions induce logarithmic oscillations in two-point functions, deviating from standard power-law scaling. At the exact location of the EP, the dilatation matrix takes a non-diagonalizable Jordan block form, and the correlator exhibits logarithmic scaling violations, characteristic of logarithmic conformal field theory (CFT) \cite{Gurarie:1993xq, Cardy:2013rqg}. 

Third, the EPs act as topological defects in the complex $N_c$ plane, generating non-Abelian geometric phases \cite{Heiss:1999qe,Berry:2004ypy, Mailybaev:2005eet, Heiss:2012dx, Wang:2021nature}. 
We show that analytic continuation along a closed contour encircling an EP induces a nontrivial monodromy, such that the identity of the physical operators is permuted upon returning to the starting point. 
This implies that the operator spectrum at the physical integer $N_c$ receives constraints from the topology of the complex $N_c$ space. 

The remainder of this paper is organized as follows.

\begin{itemize}
\item 
In Section~\ref{sec:setup}, we review basic concepts of gauge-invariant operators and introduce color-evanescent operators. 
We discuss their construction and classification of independent operator bases.

\item 
In Section~\ref{sec:calc} we detail the calculation of Gram matrices and dilatation matrices using on-shell methods.
We obtain a series of new one- and two-loop full-color results.

\item 
In Section~\ref{sec:EP-point} we analyze the $N_c$-dependence of 
the results, demonstrating the existence of negative-norm states and the emergence of complex anomalous dimensions and EPs. 

\item 
In Section~\ref{sec:physical} we discuss the physical implications of EPs,
including the $\mathcal{PT}$-symmetry phase transition, the logarithmic scaling of correlators, and the geometric phases in the complex $N_c$ plane.

\item 
In Section~\ref{sec:con} we provide a summary and discussion. The interpretation and implication of non-integer $N_c$ are addressed.
 
\item 
Technical details are provided in the appendices: the definition of operator basis in App.\,\ref{app:operators}, IR subtraction for form factors in App.\,\ref{app:IR}, symmetrized dilatation matrices in App.\,{app:symH}, $\mathcal{PT}$ and $\mathcal{C}$ matrices in App.\,\ref{app:PTmat} and \ref{app:Cmat}, the logarithmic scaling in App.\,\ref{app:Jordan}, and further discussion on geometric phases in App.\,\ref{app:geophase}.
 
 \end{itemize}

In the ancillary files, we provide the complete data of Gram and dilatation matrices, as well as their various analysis presented in the paper.

\section{Color-evanescent operators}
\label{sec:setup}

To analyze the $N_c$-dependence of the operator spectrum, we first establish an operator basis that is well-defined for arbitrary $N_c$. This leads to the central concept of \emph{color-evanescent operators}---operators that vanish for specific integer values of $N_c$ due to group identities, yet possess nontrivial norms in the analytically continued theory.

\subsection{Operator setup}

Local gauge-invariant Yang-Mills operators are constructed from the field strength
$F_{\mu\nu}=F^a_{\mu\nu}T^a$ and covariant derivatives $D_\mu$.
The generators $T^a$ of the gauge group $SU(N_c)$ satisfy the algebra $[T^a,T^b]=\mathbbm{i} f^{abc}T^c$,
and the covariant derivative acts in the adjoint representation as
\begin{equation}
\label{eq:covariantD}
D_\mu \diamond =\partial_\mu \diamond
+\mathbbm{i}g [A_\mu,\diamond] \,.
\end{equation}
A general Lorentz-scalar operator ${\cal O}(x)$ 
is formed by contracting the Lorentz and color indices of products of field strengths and their derivatives:
\begin{align}
\label{eq:YMoperator}
\mathcal{O}(x)\sim {\cal C}_{a_1,\cdots,a_L} {\cal K}(\eta)
\mathcal{W}^{a_1}_{m_1}
\mathcal{W}^{a_2}_{m_2}\cdots
\mathcal{W}^{a_L}_{m_L}\,, \\
\mbox{with}\quad
\mathcal{W}^{a_i}_{m_i}=(D_{\mu_{i_1}}\cdots
D_{\mu_{i_{m_i}}}  F_{\nu_i\rho_i})^{a_i}\,. \nonumber
\end{align}
Here, ${\cal K}(\eta)$ represents the kinematic contraction of Lorentz indices using the metric $\eta_{\mu\nu}$, and ${\cal C}_{a_1,\cdots,a_L}$ denotes the color structure formed by contracting $SU(N_c)$ indices.
The \emph{length} of an operator counts the number of field-strength insertions $\mathcal{W}^{a_i}_{m_i}$.
We restrict our attention to parity-even operators and do not consider contractions involving $\epsilon_{\mu\nu\rho\sigma}$.
For notational convenience, we denote Lorentz indices with integers; for example, the fundamental length-2 and length-3 operators are
\begin{align}
{\rm tr}(F_{\mu\nu}F^{\mu\nu})&={\rm tr}(F_{12}F_{12}), \\ 
{\rm tr}(F_{\mu_1}^{~\mu_2}F_{\mu_2}^{~\mu_3}F_{\mu_3}^{~\mu_1})&={\rm tr}(F_{12}F_{23}F_{31}).
\end{align}

Under renormalization, operators of a fixed mass dimension mix. This mixing is governed by the dilatation matrix 
$\operD$, whose eigenvalues are the anomalous dimensions $\gamma$.
By analogy with quantum mechanics, the set of operators at fixed mass dimension forms a basis of states, and the dilatation matrix acts as an effective Hamiltonian $\operH$ of the system.

To construct this Hamiltonian, one must identify a basis of independent operators modulo the equations of motion (EoM) and Bianchi identities (BI):
\begin{align}
\label{eq:EoM}
\text{EoM}: \quad  & D_\mu F^{\mu\nu} = 0 \,,  \\
\text{BI}: \quad  & D_\mu F_{\nu\rho}+D_\nu F_{\rho\mu}+D_\rho F_{\mu\nu} = 0 \,.
\end{align}

\subsection{Color-evanescent operators}

A more subtle class of relations arises from SU($N_c$) group identities that hold only for specific integer values of $N_c$. 
This naturally leads to the concept of \emph{color-evanescent operators}. We define an operator as color-evanescent at $N_c = n_0$ if it vanishes identically for this specific integer rank but remains nonzero for generic (complex) $N_c$.
While the algebraic trace relations and the resulting ``kinematic'' vanishing of states at integer $N_c$ are well known (see e.g.~\cite{Maldacena:2011jn, Shimada:2015gda, Binder:2019zqc, Caputa:2025ikn}), the central novelty of our work lies in the ``dynamic'' consequences induced by quantum corrections. Specifically, we will demonstrate how these operators alter the renormalization structure, generate complex anomalous dimensions, and drive the emergence of exceptional points.

These $N_c$-dependent identities are systematically captured using generalized Kronecker symbols.
We define a rank-$n$ Kronecker symbol as the determinant of fundamental deltas:
\begin{equation}
\label{eq:kronecker1}
\delta^{i_1\cdots i_n}_{j_1\cdots j_n}:= 
\left|
\begin{matrix}
\delta^{i_1}_{j_1} & \ldots & \delta^{i_1}_{j_n} \\
\vdots &  & \vdots\\
\delta^{i_n}_{j_1} & \ldots & \delta^{i_n}_{j_n}
\end{matrix}
\right| \,,
\end{equation}
where indices $i, j$ run from $1$ to $N_c$.

Contractions of generators with these symbols yield specific color structures. For instance, the contraction with a rank-3 symbol generates the totally symmetric tensor $d^{abc}$:
\begin{equation}
\label{eq:delta3color}
\delta^{i_1 i_2 i_3}_{j_1 j_2 j_3}T^{a_1}_{i_1 j_1}
T^{a_2}_{i_2 j_2} T^{a_3}_{i_3 j_3}
=\mathrm{tr}\big(T^{a_1}\{T^{a_2},T^{a_3}\}\big)=d^{a_1 a_2 a_3} .
\end{equation}
We refer to a color factor contracted with a rank-$n$ symbol as a $\delta_n$-color factor.
Crucially, a $\delta_n$-color factor necessarily vanishes identically for any integer $N_c < n$, since the antisymmetrization of $n$ indices is impossible in a vector space of dimension less than $n$.
However, upon analytic continuation of $N_c$ to complex values, these constraints are lifted, and the corresponding $\delta_n$ operators become independent basis elements.

Throughout this work, we utilize the sector of dimension-8, length-4 operators as a primary case study.
As detailed in Appendix~\ref{app:operators}, we classify these operators by decomposing their color structures according to the rank of the Kronecker symbols they contain. For length-4 operators, this yields a hierarchical basis:
$\{\delta_4,  \delta_3 \delta_1,  \delta_2 \delta_2\}$.
Combining these color factors with the two independent kinematic invariants,
\begin{equation}
\label{eq:dim8mono}
(F^{a}_{12} F^{b}_{23} F^{c}_{34} F^{d}_{14})\,,\quad
(F^{a}_{12}F^{b}_{12}F^{c}_{34} F^{d}_{34})\,,
\end{equation}
yields a complete basis of eight independent operators.
For computational efficiency, we further organize the basis by helicity and $C$-parity properties to block-diagonalize the dilatation matrix.
In $d=4$, the eight dimension-8 operators split into two helicity sectors: four operators in the $(-)^4$ sector and four in the $(-)^2(+)^2$ sector (see Appendix~\ref{app:operators}).

The concept of color-evanescent operators is a direct group-theoretic analogue of the evanescent operators familiar from dimensional regularization. In that context, operators are constructed using a rank-$m$ \emph{Lorentz} Kronecker delta $\delta_m$, which forces the operator to vanish when the spacetime dimension $d$ is an integer less than $m$ \cite{Buras:1989xd, Dugan:1990df, Herrlich:1994kh, Buras:1998raa, Jin:2022ivc,Jin:2022qjc}.
To distinguish these cases, we refer to the standard class as \emph{dimension}-evanescent operators and the class studied here as \emph{color}-evanescent operators. Note that an operator can be simultaneously dimension- and color-evanescent; we discuss the interplay of these effects in Section~\ref{sec:EP-point}.

The physical significance of (color) evanescent operators is substantial. When $N_c$ is treated as a continuous parameter, these operators manifest as states with indefinite norm. As we will demonstrate in Section~\ref{sec:EP-point},
this indefiniteness is the source of non-Hermiticity in the dilatation operator ({i.e.,} the effective Hamiltonian), which in turn gives rise to complex anomalous dimensions and the emergence of exceptional points. 

Our framework relies on the analytic continuation of the standard SU($N_c$) Feynman rules,  treating $N_c$ as a formal complex parameter. This treatment is consistent with the established practice of large-$N_c$ expansions and also the framework of Deligne categories (see \cite{Binder:2019zqc,Cao:2023psi}). Further discussion of this perspective is deferred to Section~\ref{sec:con}.

\section{Computational Framework: Gram and dilatation matrices}
\label{sec:calc}

To analyze the spectral properties of the operator system as a function of $N_c$, we compute two fundamental quantities: (i) the Gram matrix $G$, which defines the norm and inner product on the operator basis, and (ii) the dilatation matrix $\operD$ at  one- and two-loop orders, which governs the operator mixing and anomalous dimensions. Our calculations generalize previous single-trace results to the full-color case, retaining the exact dependence on $N_c$.

\subsection{Gram matrices}
\label{subsec:Gram}

The Gram matrix $G_{ij}$ defines the inner product on the space of operators and is extracted from the two-point correlation function:
\begin{align}
\label{eq:defineG}
\langle \symbolOa^\dag_i(x) \symbolOa_j(0)\rangle
=\frac{G_{ij}}{|x^2|^{\Delta_{\symbolOa_i}}}\,.
\end{align}
We compute $G$ at leading order in perturbation theory (zeroth order in the coupling $\alpha_s$), which is sufficient to reveal the indefinite metric structure central to our analysis.

Our computation leverages the on-shell unitarity method. The core idea is that the Gram matrix elements $G_{ij}$ can be constructed from products of tree-level form factors $\mathbf{F}_n^{(0)}$, which are the on-shell matrix elements of an operator ${\cal O}$ between the vacuum and an $n$-gluon asymptotic state (see  \cite{Yang:2019vag} for an introduction to form factors)
\begin{equation}
\mathbf{F}_{{\cal O},n} = \langle 1,...,n | {\cal O}| 0\rangle \,.
\end{equation}
By applying a unitarity cut to the two-point function---formally replacing internal propagators with on-shell delta functions---we isolate its imaginary (absorptive) part via the optical theorem, 
which factorizes into the product of two form factors:
\begin{align}
\label{eq:YMpro1}
\langle \symbolOa_L^\dag  \symbolOa_R\rangle\Big|_{\mathrm{cut}}&
=\sum_{\textrm{physical states}}
\mathbf{F}^{(0),*}_{n;\symbolOa_L} (\{p_i\}) \times
\mathbf{F}^{(0)}_{n;\symbolOa_R} (\{p_i\})\,.
\end{align}
The sum runs over all physical polarizations (helicities) and color configurations of the intermediate gluon states.
The helicity sum is performed using the standard polarization sum
\begin{align}
\label{eq:polarsum1}
\sum_{\mathrm{polarizations}}
\varepsilon_i^\mu   \varepsilon_i^\nu= \eta_{\mu\nu}-\frac{p_i^\mu \xi_i^\nu+p_i^\nu \xi_i^\mu}{p_i\cdot \xi_i}\,,
\end{align}
where $\xi_i$ is an arbitrary light-like reference momentum.
The color sum employs the SU($N_c$) completeness relation,
\begin{align}
\label{eq:lieCompelet}
\sum_a (T^a)^i_{\ j} (T^a)^k_{\ l}=\frac{1}{2}\delta^i_{l}\delta^j_{k}
-\frac{1}{2N_c}\delta^i_{j}\delta^k_{l}\,,
\end{align}
which holds for arbitrary $N_c$.

After computing this product of on-shell form factors, the full momentum-space correlator is reconstructed by reinstating the cut propagators.
The Gram matrix elements $G_{ij}$ are then extracted from the position-space correlators obtained via Fourier transform, where momenta are mapped to derivative operators acting on the position-space propagator (see \cite{Jin:2023fbz} for technical details).

\subsection{Dilatation matrices}
\label{subsec:dilatation}

The dilatation matrix $\mathbb{D}$, whose eigenvalues yield the anomalous dimensions, is extracted from the ultraviolet (UV) divergences of the form factors. Our strategy follows a three-step process: (1) compute the bare form factors using the on-shell unitarity methods, (2) isolate the UV divergences by subtracting the universal infrared (IR) divergences, and (3) extract the dilatation matrix from the remaining UV poles.

At $l$ loops, any bare form factor can be decomposed into a basis of known master integrals $I^{(l)}_i$ with coefficients $c_i$:
\begin{align}
\label{eq:FIntDecom}
\mathbf{F}^{(l)}=\sum_i c_i I^{(l)}_i\,.
\end{align}
Rather than calculating thousands of Feynman diagrams, we use modern on-shell unitarity methods, which reconstruct loop quantities from tree-level building blocks  \cite{Bern:1994zx, Bern:1994cg}. We apply the unitarity-IBP strategy \cite{Jin:2018fak, Abreu:2017xsl}: by applying generalized cuts to the loop integrand, we reduce the problem to products of tree-level amplitudes and form factors, which are then projected onto the master integral basis via integration-by-parts (IBP) reduction  \cite{Chetyrkin:1981qh,Tkachov:1981wb}:
\begin{align}
\mathbf{F}^{(l)}\Big|_{\mbox{cut}} &
= \prod (\textrm{Tree blocks})  
\rightarrow
 \sum_i  c_i \, \big( I^{(l)}_i \big|_{\mbox{cut}} \big) \,.
\end{align}

The resulting bare form factor contains IR and UV divergences, both of which appear as poles in the dimensional regulator $\epsilon = (4-D)/2$. A crucial feature of gauge theories is that the structure of IR divergences is universal and operator-independent, allowing for their unambiguous subtraction. We detail the above procedure at one- and two-loop orders below.

\subsubsection{One-loop dilatation matrices}
\label{subsec:1loopdilatation}

\begin{figure}
	\centering
	\includegraphics[width=.5\linewidth]{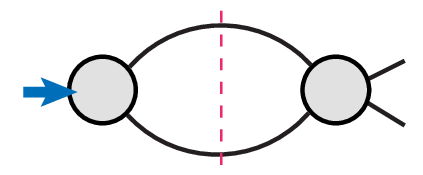}
	\caption{The one-loop cut.}
	\label{1loopcut}
\end{figure}

At one loop, the coefficients $c_i$ are determined entirely from two-particle cuts, which decompose the loop amplitude into a product of a tree-level minimal form factor and a four-gluon amplitude, as shown in Figure~\ref{1loopcut}.

For a full-color calculation, it is convenient first to consider color-decomposed tree blocks. 
The full-color results can be recovered from appropriate combinations of color-ordered tree products.
By iterating through all possible color orderings and cut channels, we reconstruct the full color dependence of the coefficients $c_i$. Further details on this procedure can be found in \cite{Jin:2022ivc}.

Once the bare form factor $\mathbf{F}^{(1)}_{\mathcal{O},B}$ is computed, the renormalized form factor $\mathbf{F}^{(1)}_{\mathcal{O},R}$ is defined by
\begin{align}
\label{eq:1loopRG}
\mathbf{F}^{(1)}_{\mathcal{O}_i,R}=\mathbf{F}^{(1)}_{\mathcal{O}_i,B}
+\sum_j \big(Z^{(1)}\big)_i^{\ j}\mathbf{F}^{(0)}_{\mathcal{O}_j,B}
\, .
\end{align}
The IR-divergent part is given by the universal formula \cite{Catani:1998bh}:
\begin{equation}
\label{eq:catani}
\mathbf{F}^{(1)}_{\mathcal{O},\mathrm{IR}}  =
\mathbf{I}_{\rm IR}^{(1)} (\epsilon)
\mathbf{F}^{(0)}_{\mathcal{O}}  \,,
\end{equation}
where $\mathbf{I}_{\rm IR}^{(1)}$ depends only on the external kinematics and color charges.

Imposing UV finiteness of the renormalized form factor determines the counterterm $Z^{(1)}$ in the modified minimal subtraction ($\overline{\rm MS}$) scheme \cite{Bardeen:1978yd}:
\begin{align}
\label{eq:1loopRG2}
0=
(\mathbf{F}^{(1)}_{\mathcal{O}_i,B}
-\mathbf{F}^{(1)}_{\mathcal{O}_i,\mathrm{IR}})\big|_{\mathrm{div}}
+\sum_j  \big(Z^{(1) }\big)_i^{\ j} \mathbf{F}^{(0)}_{\mathcal{O}_j,B}
 \,.
\end{align}

Finally, the dilatation operator $\mathbb{D}$ is defined as the logarithmic derivative of the renormalization matrix with respect to the scale $\mu$:
\begin{equation}
\label{eq:defineD}
\operD:=-\frac{d\log Z}{d\log \mu} =\sum_{l=1}^{\infty}
\Big( \frac{\alpha_s}{4\pi}
\Big)^l \operD^{(l)}\,.
\end{equation}
At one loop, this yields a simple relation between $Z^{(1)}$ and the dilatation matrix:
\begin{align}
\label{eq:Dila1}
\operD^{(1)}=2\epsilon Z^{(1)}\,.
\end{align}

\subsubsection{Two-loop dilatation matrix}
\label{subsec:dilatation2loop}

\begin{figure}[t]
	\centering
	\includegraphics[width=.87\linewidth]{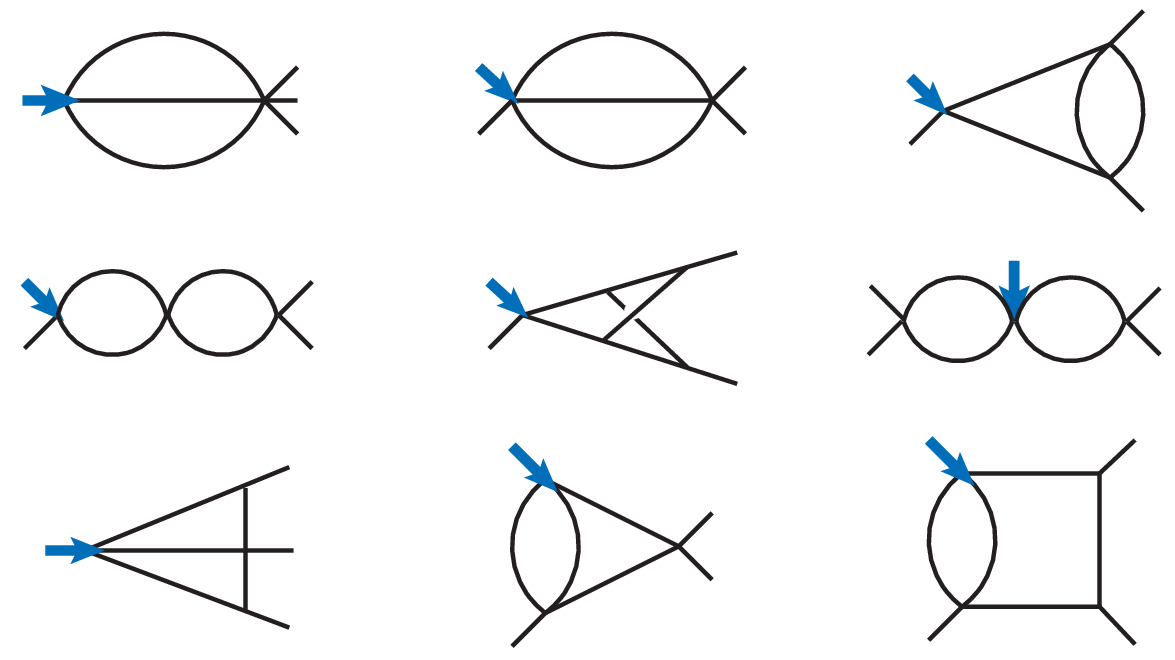}
	\caption{Two-loop master integrals. Thick blue legs represent the operator legs carrying off-shell momentum $q$.}
	\label{2loopmis}
\end{figure}

To study the exceptional points at next-to-leading order (NLO), we compute for the first time the two-loop full-color dilatation matrix for the dimension-8 operator sector. The procedure parallels the one-loop case but involves a more complex basis of master integrals (Figure~\ref{2loopmis}), which are known analytically \cite{Gehrmann:2000zt,Gehrmann:2001ck}.

\begin{figure}
	\centering
	\includegraphics[width=1\linewidth]{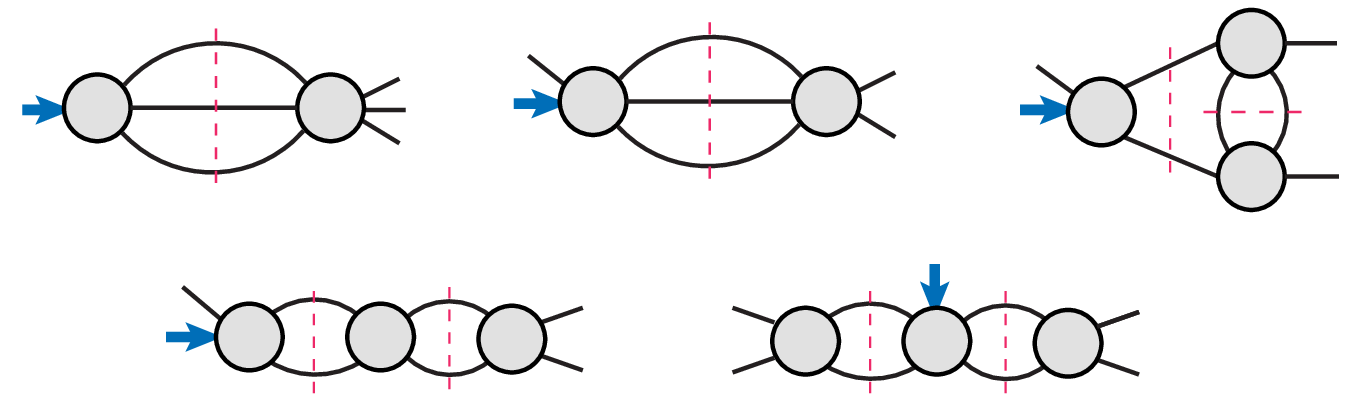}
	\caption{Cuts for computing two-loop form factors.}
	\label{2loopcuts}
\end{figure}

Determining the reduction coefficients requires evaluating the complete set of unitarity cuts shown in Figure~\ref{2loopcuts}.
The primary challenge at two loops is the combinatorial complexity of the full-color cuts. To manage this, we employ a color-decomposition strategy: for a given cut topology (e.g., the triple cut in Figure~\ref{2loopcuts}), we fix the color ordering on one side of the cut and sum over the permutations of color-ordered diagrams on the other side. This procedure reconstructs the full-color integrand, which is subsequently reduced to master integrals using the FIRE6 package \cite{Smirnov:2019qkx}.

After computing the bare form factors, we isolate the UV divergence by subtracting the universal IR poles. At two loops, the IR structure is given by \cite{Catani:1998bh, Aybat:2006mz}:
\begin{align}
\mathbf{F}^{(2)}_{\mathcal{O},\mathrm{IR}}  =
\mathbf{I}_{\rm IR}^{(2)} (\epsilon)
\mathbf{F}^{(0)}_{\mathcal{O}} +\left(\mathbf{I}_{\rm IR}^{(1)} (\epsilon)
\mathbf{F}^{(1)}_{\mathcal{O},R}\right)\bigg|_{\mathrm{div}}\,, 
\label{eq:catani2loop}
\end{align}
where the two-loop IR operator $\mathbf{I}_{\rm IR}^{(2)}$ contains universal kinematic and color-charge-dependent terms; see Appendix~\ref{app:IR} for details.

The two-loop dilatation matrix $\operD^{(2)}$ is then extracted from the renormalization matrix $Z^{(2)}$ via 
\begin{align}
	\operD^{(2)}=4\epsilon Z^{(2)}-2\epsilon \left(Z^{(1)}\right)^2+2\beta_0Z^{(1)}\,,
	\label{gamma2}
\end{align}
with $\beta_0$ being the one-loop beta function coefficient.

We emphasize that our results have passed several rigorous consistency checks. 
First, coefficients of master integrals determined from multiple distinct cuts are mutually consistent. 
Second, the IR divergences match the universal Catani formula, and the $1/\epsilon^2$ UV poles in $Z^{(2)}$ cancel precisely with $(Z^{(1)})^2$ as required by the renormalization group equation according to Eq.\,\eqref{gamma2}.
Third, the dilatation matrices exhibit the correct block-diagonal structure dictated by the basis choice.
Finally, and most nontrivially, the product of the dilatation matrix and the Gram matrix, $\mathcal{D}^{(1)}\cdot G^{(0)}$ is real and symmetric, confirming the internal consistency of our operator basis and inner-product definitions.

\section{Non-hermiticity and exceptional points}
\label{sec:EP-point}

In this section, we present the main computational results, establishing a direct link between the analytic structure of Yang-Mills theory and the physics of non-Hermitian systems.
First, we show that color-evanescent operators---which vanish identically at specific integer values of $N_c$---become negative-norm states once $N_c$ is analytically continued away from those integers. This renders the Gram matrix, which defines the metric on the operator space, indefinite.
Second, we show that this indefinite metric induces non-Hermiticity in the dilatation operator, leading to the emergence of complex anomalous dimensions (ADs) and exceptional points (EPs) in the operator spectrum.
Third, we explore higher-dimensional operators, where the interplay between color- and dimension-evanescent operators generates an even richer spectral structure.
 
\subsection{Negative-norm states}
\label{sec:negative-norm}

We begin by demonstrating the connection between color-evanescent operators and the indefinite metric, using the $(-)^4$ helicity sector of dimension-8 operators as our case study (see Eq.\,\eqref{eq:newbaselen4}).

The corresponding leading-order Gram matrix, defined in Eq.\,(\ref{eq:defineG}), is given by:
\begin{small}
\begin{align}
\label{eq:Galpha}
&G^{(0)}=\frac{6144(n^2-1)}{n}\times
\\
&
\begin{pmatrix}
 \frac{15 a b u}{2 n} & -\frac{5 a b u}{n} & -10 a b c & -10 a b c \\
*  & \frac{2 a \left(13 n^3+2 n^2-25 n-30\right)}{n} & 4 a c (3 n-5) & -4 a c (n+5) \\
 * & * & 8 n \left(2 n^2+1\right) & 8 n \left(n^2+3\right) \\
 * & * & * & 8 n \left(3 n^2-1\right) 
\end{pmatrix}
,\nonumber
\end{align}
\end{small}
where $n:=N_c$, and
\begin{align}
\label{eq:notation1}
& a:=N_c-2 \,, \qquad b:=N_c-3\,, \nonumber\\
& c:=N_c+1\,, \qquad u:=3N_c^2+7N_c+6 \,.
\end{align}
The lower triangle is omitted as $G^{(0)}$ is symmetric. 

The structure of $G^{(0)}$ directly reflects the evanescent nature of the basis operators. 
The first row and column of Eq.\,(\ref{eq:Galpha}) vanish at $N_c=3$, 
consistent with the fact that $\mathcal{O}_1$ in Eq.\,(\ref{eq:newbaselen4}) 
is color-evanescent at $N_c=3$.
Similarly, the first two rows and columns  vanish at $N_c=2$ because both
$\mathcal{O}_{1}$ and $\mathcal{O}_2$ are color-evanescent at $N_c=2$.

The signature of  $G^{(0)}$ (i.e., the signs of its eigenvalues), summarized in 
Table~\ref{tab:nearinteger8a}, reveals the presence of negative-norm states for $N_c<3$.
In general, a $\delta_{(n+1)}$ operator $O_a$ vanishes for integer $N_c$ smaller than $n+1$; consequently, the matrix elements $G_{ai}$ contain factors of the form $(N_c-1)\cdots (N_c-n)$, causing the norm to flip sign as $N_c$ traverses the integers $1,2,\cdots,n$.
For length-4 operators, the rank of the Kronecker symbol in color space
is at most 4, ensuring that all operators possess positive norms for $N_c>3$.

\begin{table}
\renewcommand\arraystretch{1.2}
\centering
\vspace{0.4cm}
\begin{tabular}{|c|c|c|c|c|c|}
\hline
 $0<N_c<1$ & $1<N_c<2$ & $2<N_c<3$ & $3<N_c$\\
\hline
 $(-,+,-,-)$ & $(+,-,+,+)$ & $(-,+,+,+)$ & $(+,+,+,+)$\\
\hline
\end{tabular}
\caption{\label{tab:nearinteger8a} The signature of the Gram matrix in
Eq.\,\eqref{eq:Galpha}. }
\end{table} 

For $N_c<3$, the Gram matrix is indefinite. In the analytically continued theory, this signals the loss of positive-definite inner product and hence the breakdown of unitarity. It is precisely this feature that permits the appearance of complex anomalous dimensions.

\begin{figure*}
\centering
\subfigure[]
{\includegraphics[width=0.4\linewidth]{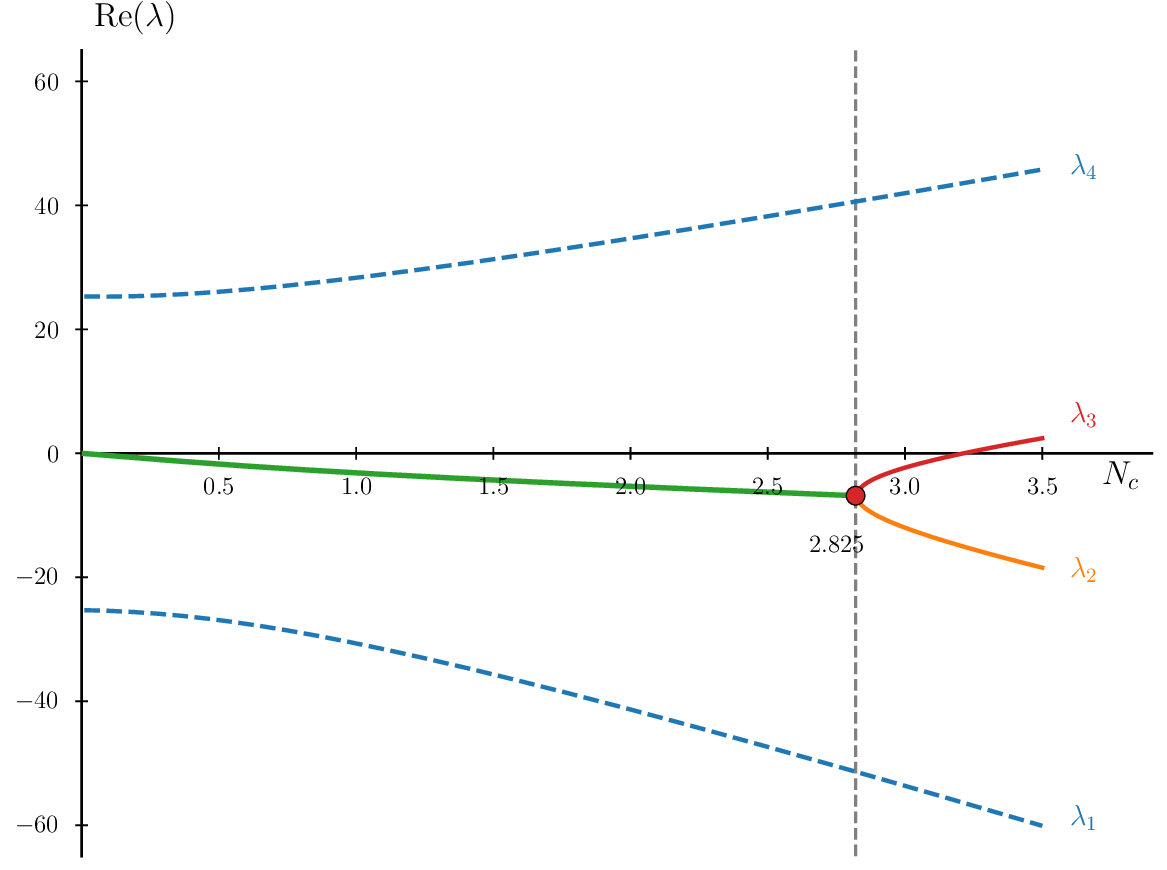}}
\hspace{0.9cm}
\subfigure[]
{\includegraphics[width=0.4\linewidth]{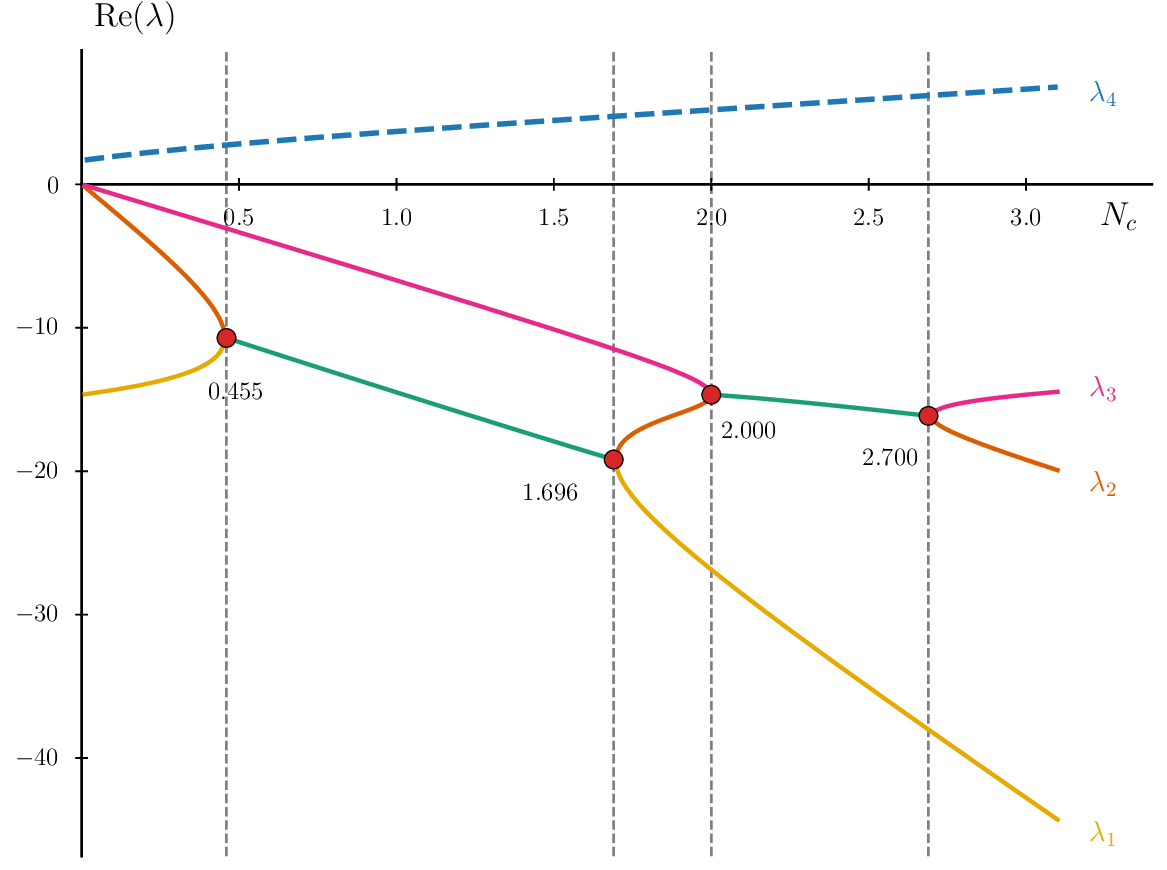}}
\caption{\label{fig:d8L4sec}The one-loop spectrum of dim-8 length-4 sector. 
  The green line denotes a complex conjugate pair.
  (a) The $(-)^4$ sector. (b) The $(-)^2(+)^2$ sector. }
\end{figure*} 

To make this explicit, one can diagonalize the real symmetric Gram matrix. While a standard real congruence transformation yields the signature matrix ${\cal P}$, we define a transformation matrix $M$ that absorbs the negative signs of the signature matrix, so that the transformed metric becomes the identity (see Appendix~\ref{app:symH} for more details):
\begin{equation}
\label{eq:Mproperty0}
M\cdot G\cdot M^{\mathrm{T}}=\mathbbm{1}\,,\quad
M^*\cdot G\cdot M^{\mathrm{T}}=\mathcal{P}\,.
\end{equation}
Here $\mathcal{P}$ is a diagonal matrix  whose entries are the signs of the eigenvalues of $G$. Note that for an indefinite signature, the condition $M\cdot G\cdot M^{\mathrm{T}}=\mathbbm{1}$ renders the transformation matrix $M$ necessarily complex.
Using the transformation matrix $M$, we define the orthonormal basis:
\begin{equation}
\label{eq:defineOp1}
{\cal O}^{\mathrm{p}}_i:=M_{ij} \, {\cal O}_j\,.
\end{equation}
On this basis, the two-point functions are diagonal and carry the sign of the metric:
\begin{equation}
\label{eq:defineP}
\langle {\cal O}^{\mathrm{p}\dag}_i (x) \, {\cal O}^{\mathrm{p}}_j(0)\rangle=
\frac{\mathcal{P}_{ij}}{|x^2|^{\Delta_i}}\,.
\end{equation}

\subsection{Exceptional points in spectrum}
\label{sec:fullcolordim8}

We now demonstrate how the indefinite metric leads to the emergence of exceptional points. The one-loop dilatation matrix $\mathbb{D}^{(1)}$ for the above dimension-8, length-4 sector discussed above is given by:
\begin{equation}
\label{eq:oldDila1}
\mathbb{D}^{(1)}=
2
\scalebox{0.81}{
$\displaystyle
\begin{pmatrix}
\frac{2(15-8N_c)}{3} & 5(3-N_c) & 0 & 0
\\[0.2em]
2(N_c-4) & \frac{11N_c-36}{3} & 2(N_c-2) & 2(N_c-2)
\\[0.2em]
-14 & -21 & \frac{14(N_c+3)}{3} & -2(2N_c+7)
\\[0.2em]
-12 & -18 & 12 & -\frac{2(11N_c+18)}{3}
\end{pmatrix}
$}.
\end{equation}
Its spectrum of anomalous dimensions, plotted in Figure~\ref{fig:d8L4sec}(a), exhibits a characteristic non-Hermitian transition
at $N_c^{\rm ep}=2.82466$: two real eigenvalues coalesce and subsequently bifurcate into a complex conjugate pair.

The location of the transition point can be found by computing the characteristic polynomial of $\mathbb{D}^{(1)}$ and extracting its discriminant:
\begin{align}
\label{eq:discrim1}
\bar\Delta  =
N_c^2 & \big(5904 N_c^{10}+504576 N_c^8+5930561 N_c^6
\\
&-64346120 N_c^4-135238400 N_c^2-72704000 \big)\,,\nonumber
\end{align}
whose only positive real root is $N_c^{\rm ep}=2.82466$.%
\footnote{The overall factor $N_c^2$ in Eq.\,(\ref{eq:discrim1}) arises from the rescaling $\alpha_s\rightarrow \alpha_s N_c$ and  does not indicate an EP at $N_c=0$.
Similarly, an overall negative power of $N_c$ does not imply an EP at $N_c=\infty$.
As discussed in Section~\ref{sec:geemetric}, the points $N_c=0, \infty$ are distinguished from EPs by the absence of a nontrivial geometric phase.}

\begin{figure}
\centering
\includegraphics[width=0.7\linewidth]{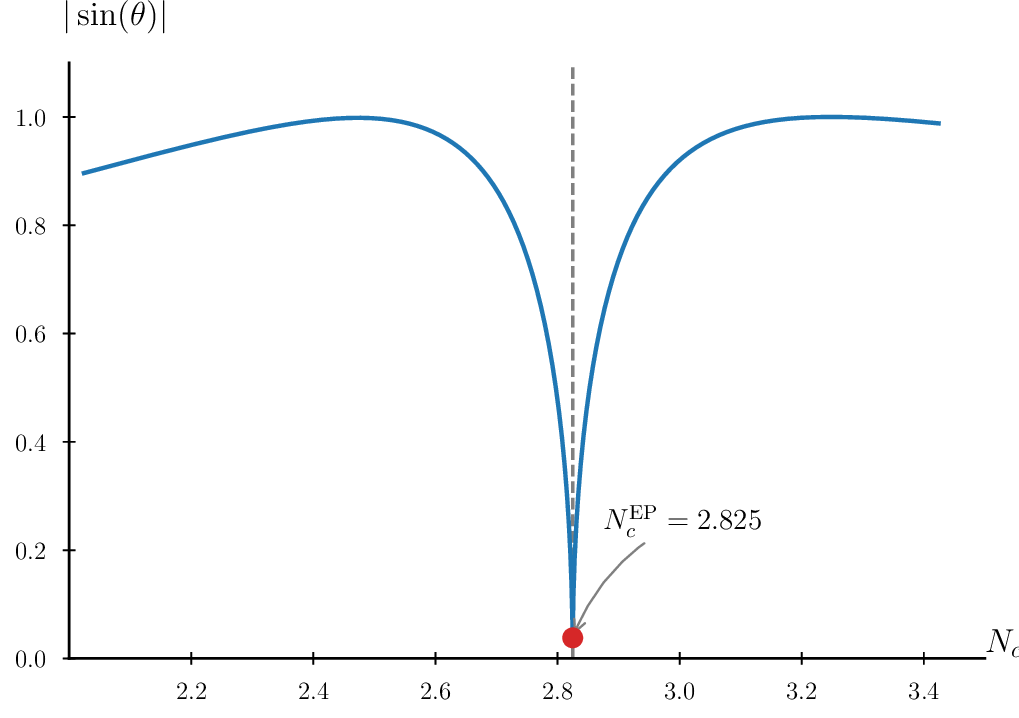}
\caption{The angle $\theta$ between the two eigenvectors of Eq.\,(\ref{eq:oldDila1})
which coalesce at  $N_c^{\mathrm{EP}}=2.825$.}
\label{fig:angle}
\end{figure}

The corresponding two eigenstates also coalesce, as illustrated in Figure~\ref{fig:angle}, where the angle between the two eigenvectors vanishes precisely at $N_c^{\rm ep}$.
Here the angle $\theta$  is defined through
$\cos\theta=|\langle v_2| v_3 \rangle|/\sqrt{\langle v_2| v_2 \rangle
\langle v_3 | v_3 \rangle}$, where $\langle v_i|v_j \rangle=v_i^* \cdot v_j $ is the Euclidean norm.
As shown, the angle vanishes precisely at $N_c^{\rm ep}$, confirming the coalescence (parallel alignment) of the two eigenvectors.
The simultaneous coalescence of eigenvalues and eigenvectors is the defining hallmark of an exceptional point (EP) \cite{kato2013perturbation}.
%

The above phenomenon is a universal feature in general operator sectors.
The $(+)^2(-)^2$ sector at length-4 dimension-8 contains four operators, whose definitions are given in Eq.\,(\ref{eq:newbaselen4b}). 
Their one-loop spectrum exhibits multiple real, positive EPs, resulting in
a richer structure along the positive real $N_c$-axis, as shown in Figure~\ref{fig:d8L4sec}(b).

\begin{figure*}
\centering
\subfigure[]
{\includegraphics[width=0.4\linewidth]{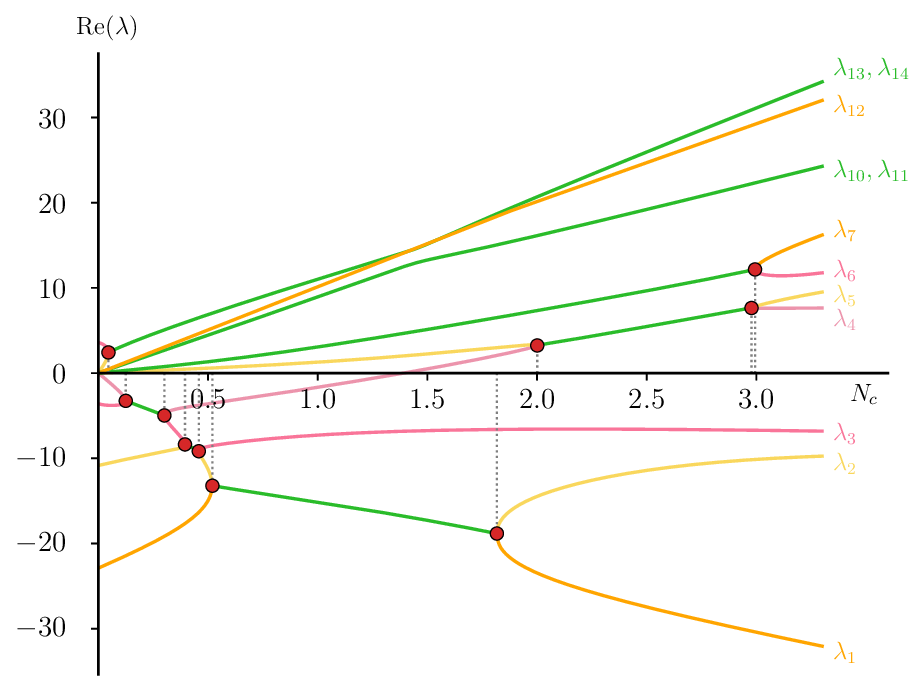}}
\hspace{0.9cm}
\subfigure[]
{\includegraphics[width=0.4\linewidth]{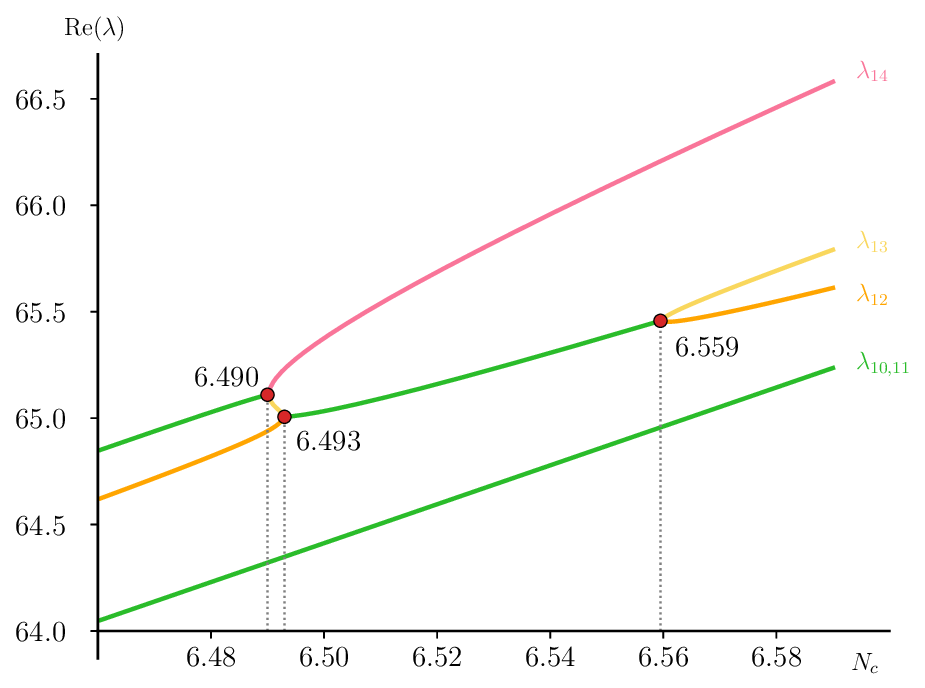}}
\caption{\label{D12h4d2-b}The real spectrum in the dimension-evanescent $D$-(2,2) sector of dimension-12 length-4.
The green lines refer to  complex conjugate pairs.
Four ADs $\lambda_{8,9,15,16}$ are not shown because they are always real. (a) $N_c<4$ range. (b) $N_c>6$ range.}
\end{figure*} 

Starting from the two-loop level, the $(-)^4$ and $(-)^2(+)^2$ sectors mix, requiring the diagonalization of the full eight-operator system.
Our calculation confirms that these EPs persist at NLO, with their locations receiving perturbative corrections.
For instance, the EP at $N_c=2.825$ is shifted by $-4.518 \frac{\alpha_s}{4\pi}$.
The two-loop corrections to the positive and real EPs shown in Figure~\ref{fig:d8L4sec} 
are given as follows
\begin{align}
 \{&2.825-4.518 \frac{\alpha_s}{4\pi},\,
  0.455-5.407 \frac{\alpha_s}{4\pi},\,
  1.696+15.42 \frac{\alpha_s}{4\pi},\,
  \nonumber\\
  & 2+3.865 \frac{\alpha_s}{4\pi},\,
  2.700+18.72 \frac{\alpha_s}{4\pi}\}\,.
\end{align}
Note that the one-loop EP located precisely at $N_c=2$ is shifted at NLO, demonstrating that its coincidence with an integer value is accidental rather than symmetry-protected. 
In contrast, the locations where the signature of the Gram matrix flips are fixed at exact integers by the color algebra and are robust against loop corrections.

The origin of the EP lies in the non-Hermiticity of the dilatation operator, which in turn stems from the indefinite Gram metric.
This can be made manifest by constructing the \emph{symmetrized dilatation matrix} $\operH$:
\begin{align}
\label{eq:symH}
\mathbb{H}:=M\cdot \mathbb{D}\cdot M^{-1}
=M\cdot \mathbb{D} \cdot G \cdot M^{\mathrm{T}}\,,
\end{align} 
where $M$ is the transformation that diagonalizes the Gram matrix as introduced in Eq.\,(\ref{eq:Mproperty0}).
As detailed in Appendix~\ref{app:symH}, $\mathbb{H}$ is Hermitian if and only if the metric $G$ is positive-definite.
For indefinite $G$, however, the transformation matrix $M$ becomes complex, and the resulting $\mathbb{H}$ is generically non-Hermitian.

For the sector in Eq.\,(\ref{eq:newbaselen4}), $G$ is positive-definite for $N_c>3$, 
rendering $\mathbb{H}$ Hermitian and guaranteeing a real spectrum, consistent with unitarity. For $N_c<3$, $G$ is indefinite, $\mathbb{H}$ becomes non-Hermitian and permits complex eigenvalues. 
The particularly interesting interval is $(N_c^{\rm ep}, 3)$, where the spectrum remains entirely real despite the non-Hermiticity of $\mathbb{H}$. This signals a phase of unbroken $\mathcal{PT}$ symmetry, which we analyze in Section~\ref{sec:physical}.

\subsection{Interplay of color- and dimension-evanescent operators}
\label{sec:TwoEvaTypes}

We have seen above that color-evanescent operators contribute to negative-norm states.  
Meanwhile, the dimension-evanescent operators can also give rise to negative-norm states at large $N_c$ for $d=4-2\epsilon$ ($\epsilon<0$) \cite{Jin:2023cce, Jin:2023fbz} (see also \cite{Hogervorst:2015akt, Ji:2018yaf} for scalar and fermion theories).
Although these operators vanish at strictly integer dimension, their analytic continuation is well-defined within dimensional regularization and influences the structure of operator mixing in the extended parameter space.
In this subsection, we explore the rich phenomenology that arises from the interplay of color- and dimension-evanescent operators.
The striking feature is that operator sectors containing dimension-evanescent operators can exhibit complex anomalous dimensions and EPs even at large $N_c$.

As discussed previously, the color factors of length-4 operators do not generate negative norms for $N_c>3$. Consequently, a pure length-4 sector without dimension-evanescent operators would possess a real spectrum in this regime. However,  if such a sector contains dimension-evanescent operators, the metric can remain indefinite for $N_c>3$, leading to exceptional points at high $N_c$.

We illustrate this with a concrete example: a sector of dimension-evanescent operators at mass dimension 12, 
denoted as $D$-$(2,2)$ sector (comprising operators of the form $\partial_\mu\partial_\nu {\cal O}^{\mu\nu}$, i.e., second derivatives of rank-2 tensor). 
The operators in this basis are evanescent in both color and spacetime dimension, 
with norms that vanish for specific integer values of $N_c$ and $d$.
Together, we classify these operators by the rank of their Kronecker symbols in both color ($n$) and Lorentz ($m$) space, denoted as ${\cal O}^{n,m}_i$.
The polynomial factors governing their norms are summarized below:
\begin{align}
\label{eq:size16dim12}
&\mathcal{O}^{2,6}_1:\  \quad (d-4)(d-5)(N_c-1)\,,  \nonumber\\
&\mathcal{O}^{2,5}_{1,\cdots,7}:\  (d-4)(N_c-1)\,, 
 \\
&\mathcal{O}^{3,6}_1:\  \quad (d-4)(d-5)(N_c-1)(N_c-2)\,, 
\nonumber \\
&\mathcal{O}^{3,5}_{1,\cdots,4}:\  (d-4)(N_c-1)(N_c-2)\,, 
\nonumber \\
&\mathcal{O}^{4,5}_{1,\cdots,3}:\  (d-4)(N_c-1)(N_c-2)(N_c-3)\,.
\nonumber 
\end{align}
All 16 operators in this sector vanish identically at $d=4$, but define a nontrivial analytic sector for $d\neq4$. Furthermore, two of them, $\mathcal{O}^{2,6}_1$ and $\mathcal{O}^{3,6}_1$, contain $\delta_6$ Lorentz Kronecker symbols and vanish at $d=5$ as well.
The signature of the Gram matrix for $4<d<5$ is:
\begin{equation}
\begin{tabular}{|c|c|c|c|c|}
\hline
  & $0<N_c<1$ & $1<N_c<2$ & $2<N_c<3$ & $3<N_c$  \\
\hline
 $(n_{+},n_{-})$
& $(5,11)$ & $(11,5)$ & $(11,5)$ & $(14,2)$ \\
\hline
\end{tabular}
\end{equation}
Crucially, the presence of the $\delta_6$ Lorentz structure guarantees two negative-norm states even for $N_c>3$, rendering the metric indefinite at arbitrarily large color number.

\begin{figure}
  \centering
{
\includegraphics[width=1\linewidth]{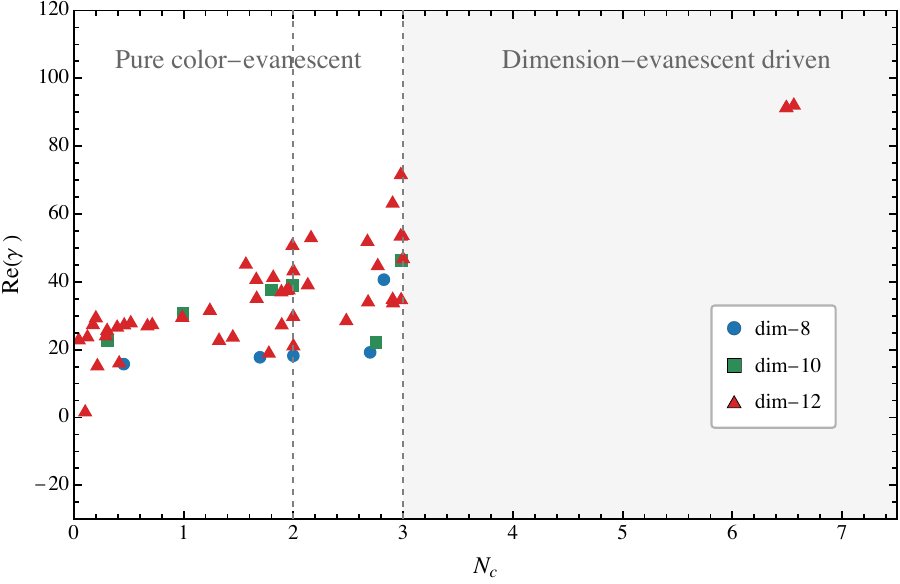}
}
\caption{\label{fig:multi} The distribution of real and positive EPs for length-4 operators with dimension 8,10,12.
EPs with $N_c>3$ are driven by dimension-evanescent operators.} 
 \end{figure} 

In Figure~\ref{D12h4d2-b} we plot the real parts of ADs in this sector.
Panel (b) reveals three EPs and branches of complex anomalous dimensions occurring at $N_c>3$,
a phenomenon absent in length-4 sectors without dimension-evanescent operators.
Notably, even beyond the largest EP $N_c^{\mathrm{max}}=6.559$,
a pair of complex conjugate anomalous dimensions persists, directly tied to the two negative-norm operators $\mathcal{O}_1^{2,6}$ and $\mathcal{O}_1^{3,6}$.

Finally, Figure~\ref{fig:multi} summarizes the distribution of real, positive EPs for length-4 operators up to mass-dimension 12.
All three EPs with $N_c>3$ originate exclusively from sectors containing $\delta_6$ dimension-evanescent operators.
In their absence, the EPs of length-4 operators are strictly confined to the region $N_c<3$. Similarly, for length-5 sectors, EPs appear at $N_c<4$ but can go beyond this bound with dimension-evanescent operators (see Appendix~\ref{app:operators} and ancillary files).
This demonstrates that analytic continuation in spacetime dimension and in color number act as independent sources of metric indefiniteness, whose combined effect produces a qualitatively richer non-Hermitian structure.

\section{Physical effects of exceptional points}
\label{sec:physical}

Although the number of colors $N_c$ takes only positive integer values in the physical world, its analytic continuation reveals nontrivial structural and physical consequences.
In this section, we explore three key physical implications.
First, we show that the existence of EPs signals a phase transition
characterized by the spontaneous breaking of an effective $\mathcal{PT}$ symmetry.
Crucially, we show that this emergent symmetry of the dilatation matrix is directly induced by the fundamental spacetime $\mathcal{PT}$ symmetry of the underlying YM theory.
Second, we show that the transition from real to complex spectra across an EP induces logarithmic oscillations and scaling behavior in two-point correlators.
Third, we establish that analytic continuation along paths encircling EPs generates non-Abelian geometric phases, uncovering hidden topological structures in the complex $N_c$ space.

\subsection{EP and PT-symmetry}
\label{sec:PT}

As discussed in Section~\ref{sec:fullcolordim8}, 
the symmetrized dilatation matrix for the length-4 dimension-8 operators is Hermitian for $N_c>3$ 
but becomes non-Hermitian for $N_c<3$. 
The non-Hermitian region is further divided into two distinct phases:
an interval $(N_c^{\rm ep},3)$ where the spectrum remains real, and an interval $(0, N_c^{\rm ep})$ where complex conjugate eigenvalues emerge. 
Such a transition from a real to a complex spectrum is the hallmark of a system possessing an underlying Parity-Time ($\mathcal{PT}$) symmetry \cite{Bender:2003gu,Bender:2023cem}.

To understand the physical origin of this symmetry, recall that the dilatation matrix $\mathbb{D}$ acts as the Hamiltonian of operator mixing.
If an operator ${\cal O}=v_i \mathcal{O}_i$ is an eigen-operator, 
the coefficients $v_i$ form a left eigenvector of $\mathbb{D}$. 
Through the similarity transformation $M$ defined in Eq.\,(\ref{eq:Mproperty0}), we map to the orthonormal basis ${\cal O}_i^{\mathrm{p}}$ where the effective Hamiltonian is the symmetrized matrix ${\operH}$. 
Let $v_h$ denote an eigenstate vector in this symmetrized representation.

We define the effective time-reversal operation, $T_{\mathrm{eff}}$, 
as the anti-linear complex conjugation operation:
\begin{align}
\label{eq:Tope}
T_{\mathrm{eff}}\circ v_h=v_h^*\,,\quad
T_{\mathrm{eff}}\, \mathbb{H} \, T_{\mathrm{eff}}^{-1}=
\mathbb{H}^*\,.
\end{align}
The crucial insight is that the effective parity operation, $P_{\mathrm{eff}}$, 
is realized by the signature matrix of the operator-space metric. 
Since the Gram matrix is diagonal in the orthonormal basis with entries $\mathcal{P}_{ij}= \pm1$ (see Eq.\,(\ref{eq:defineOp1})), 
we identify the parity operator with the signature matrix itself: $P_{\mathrm{eff}}=\mathcal{P}$.
The composite $\mathcal{PT}$ operation acting on a state $v_h$ is then
\begin{align}
\label{eq:PToverCoe0}
P_{\mathrm{eff}}T_{\mathrm{eff}}\circ v_h  =\mathcal{P}\cdot v_h^* \,,
\end{align}

By construction, these operations satisfy the algebra of a $\mathcal{PT}$-symmetric system:
\begin{align}
P_{\mathrm{eff}}^2=\mathbbm{1}\,, \quad
[P_{\mathrm{eff}},T_{\mathrm{eff}}]=0\,,
\end{align} 
and, importantly, the Hamiltonian satisfies $[P_{\mathrm{eff}} T_{\mathrm{eff}},\mathbb{H}]=0$,
i.e., it is $\mathcal{PT}$-symmetric.

The consequences of this symmetry are well-established \cite{Bender:2003gu,Bender:2023cem}. The region $(N_c^{\rm ep},3)$ corresponds to the $\mathcal{PT}$-unbroken phase, where the eigenvectors of $\operH$ are simultaneous eigenstates of $P_{\mathrm{eff}} T_{\mathrm{eff}}$, guaranteeing a real spectrum. The EP marks the transition point at which this symmetry becomes spontaneously broken; in the broken phase $(0,N_c^{\rm ep})$, the eigenstates are no longer invariant, and eigenvalues bifurcate into complex conjugate pairs.

A key feature of $\mathcal{PT}$-symmetry is that, in the $\mathcal{PT}$-unbroken phase, the reality of the spectrum allows one to redefine the inner product (the $\mathcal{CPT}$ inner product) to restore positive norm of states \cite{Bender:2002vv}. 
This is achieved by constructing a $\mathcal{C}$ matrix which is used to defined a $\mathcal{C}$-conjugated operator basis 
(${\cal O}^{\mathrm{p}}_i$ is defined in Eq.\,(\ref{eq:defineOp1})):
\begin{align}
{\cal O}^{\mathrm{cp}}_i:=  {\cal O}^{\mathrm{p}}_j\mathcal{C}_{ji}\,,
\end{align}
and the new Gram matrix $G^{\mathrm{c}}_{ij}$ is positive definite in the $\mathcal{PT}$-unbroken phase
\begin{align}
\langle {\cal O}^{\mathrm{p}\dag}_i(x) {\cal O}^{\mathrm{cp}}_j(0)\rangle
=\frac{(\mathcal{P}\mathcal{C})_{ij}}{|x^2|^{\Delta_i}}
=:\frac{G^{\mathrm{c}}_{ij}}{|x^2|^{\Delta_i}}\,.
\end{align}
The detailed construction of the $\mathcal{C}$ matrix and its implications are discussed in 
Appendix~\ref{app:Cmat}.

To elucidate the physical meaning of this effective $\mathcal{PT}$-symmetry, we transform the $\mathcal{PT}$ operators back to the basis of the original dilatation matrix $\mathbb{D}$. 
The induced operations on the coefficient vector $v$,
denoted by  $\tilde{T}_{\mathrm{eff}}$ and $\tilde{P}_{\mathrm{eff}}$,
are derived from the similarity transformation relating $\mathbb{H}$ and $\mathbb{D}$:
\begin{equation}
\begin{aligned}
\label{eq:newPT}
&\tilde{P}_{\mathrm{eff}}\circ v 
=M^{\mathrm{T}}\cdot \mathcal{P}\cdot (M^{\mathrm{T}})^{-1}\cdot v\,, \\
&\tilde{T}_{\mathrm{eff}}\circ v 
=M^{\mathrm{T}}\cdot ( (M^{\mathrm{T}})^{-1}\cdot v)^*\,.
\end{aligned}
\end{equation}
On this original operator basis, the composite operation simplifies to pure complex conjugation:
\begin{align}
\label{eq:PToverCoe}
\tilde{P}_{\mathrm{eff}}\tilde{T}_{\mathrm{eff}}\circ v = v^* \,.
\end{align}

\begin{table}[!t]
\renewcommand\arraystretch{1.3}
\centering
\begin{tabular}{|c|c|c|c|} 
\hline
 & $\hat{P}$ & $\hat{T}$ & $\hat{P}\hat{T}$\\
\hline
$\mathbbm{i}$ & $+$ & $-$ & $-$\\
\hline
$\partial_\mu$ & $(-)_\mu$ & $-(-)_\mu$ & $-$\\
\hline
$A_\mu$ & $(-)_\mu$ & $(-)_\mu$ & $+$\\
\hline
\end{tabular}
\caption{\label{tab:PT0}
Transformation of gluon fields and derivatives under spacetime $\mathcal{PT}$-symmetries, where $(-)_0=1$, $(-)_i=-1$.}
\end{table}

This action has a direct correspondence to the spacetime $\mathcal{PT}$ transformation of the gauge field operators.
Denote the spacetime $\mathcal{PT}$ operator as $\hat{P}$ and $\hat{T}$.
The transformation properties of the elementary gauge field and derivative are summarized in Table~\ref{tab:PT0}.
The combination of spacetime parity and time-reversal acts on a composite operator ${\cal O}=v_i \mathcal{O}_i$ as 
\begin{equation}
\label{eq:PToverV}
\hat{P}\hat{T}\circ\big( v_i \mathcal{O}_i(x^\mu)\big)
=v_i^*\big(\hat{P}\hat{T}\circ\mathcal{O}_i(x^\mu)\big)
= v_i^* \mathcal{O}_i(-x^\mu)\,.
\end{equation}
Comparing Eq.\,(\ref{eq:PToverCoe}) and Eq.\,(\ref{eq:PToverV}), we identify:
\begin{equation}
\tilde{P}_{\mathrm{eff}}\tilde{T}_{\mathrm{eff}}\simeq (\hat{P}\hat{T})_{\rm spacetime} \,.
\end{equation} 
Thus, the emergent $\mathcal{PT}$ symmetry is precisely the manifestation of spacetime $\mathcal{PT}$ symmetry acting on the operator algebra. The $\mathcal{PT}$-unbroken phase is a regime where spacetime $\mathcal{PT}$ symmetry protects the reality of the anomalous dimension spectrum. 
The EPs mark the spontaneous breaking of the $\mathcal{PT}$ symmetry, providing a fundamental field-theoretic interpretation of the non-Hermitian spectrum.
More details are given in Appendix~\ref{app:PTmat}.

\subsection{Logarithmic behavior of two-point functions}
 \label{sec:log}

The emergence of complex anomalous dimensions in the $\mathcal{PT}$-broken phase has an observable consequence: it fundamentally alters the scaling behavior of two-point correlation functions, inducing deviations from the conventional power-law decay of unitary theories.

Without loss of generality, we analyze the two-point functions of the eigen-operators ${\cal O}_\pm$ associated with the eigenvalues $\lambda_\pm$ that coalesce at the EP.
The structure of these correlators depends critically on the phase of the system.

In the $\mathcal{PT}$-unbroken phase $(N_c^{\rm ep}< N_c <3)$, 
the anomalous dimensions $\lambda_\pm$ are real.
The eigen-operators are orthogonal with respect to the Gram matrix ($G^{>}_{+-}=G^{>}_{-+}= 0$),
and the diagonal Green functions $\mathcal{G}^{>}_{++}(x)$ and $\mathcal{G}^{>}_{--}(x)$ take the standard power-law form:
\begin{align}
\mathcal{G}^{>}_{++}(x)=\frac{G^{>}_{++}}{|x^2|^{\Delta_{{\cal O}}+\lambda^{>}_{+}}}\,,
\quad
\mathcal{G}^{>}_{--}(x)=\frac{G^{>}_{--}}{|x^2|^{\Delta_{{\cal O}}+\lambda^{>}_{-}}}\,.
\end{align}

In the $\mathcal{PT}$-broken phase $(0< N_c <N_c^{\rm ep})$, the situation changes dramatically. 
The anomalous dimensions become a complex conjugate pair,
\begin{equation}
\lambda^{<}_{\pm}=a\pm\mathbbm{i}\omega \,.
\end{equation}
Crucially, the eigen-operators are no longer orthogonal; instead, they become null vectors with vanishing norm,
\begin{equation}
G^{<}_{++}=G^{<}_{--}=0\,.
\end{equation}
However, the off-diagonal metric elements are non-zero, leading to cross-correlators:
\begin{align}
\mathcal{G}^{<}_{+-}(x)=\frac{G^{<}_{+-}}{|x^2|^{\Delta_{{\cal O}}+\lambda^{<}_{-}}}\,,
\quad
\mathcal{G}^{<}_{-+}(x)=\frac{G^{<}_{-+}}{|x^2|^{\Delta_{{\cal O}}+\lambda^{<}_{+}}}\,,
\end{align}
where $\lambda^{<}_{-}=(\lambda_{+}^{<})^*$ and $G^{<}_{+-}=(G_{-+}^{<})^*$.
These functions are complex-valued.

\begin{figure}
\centering
\includegraphics[width=0.99\linewidth]{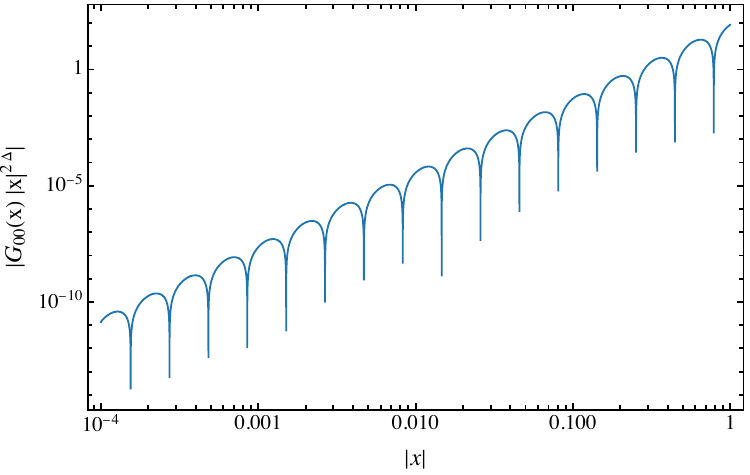}
\caption{\label{fig:oscill}The ``LogLogPlot" for the distance dependence of $|\mathcal{G}_{00}^{<}(x)|$  in Eq.\,(\ref{eq:oscillateG}), 
evaluated at $N_c=1.7$ ($\mathcal{PT}$-broken phase), $\alpha_s=1/3$, $\Delta_{O}=8$. 
The coupling $\alpha_s$ enters the AD $a\pm\mathbbm{i} \omega$ as an overall factor. }
\end{figure}

To construct a real physical observable, we consider a linear combination of the eigen-operators:
\begin{equation}\label{eq:O0}
{\cal O}_0:={\cal O}_- - {\cal O}_+ \,.
\end{equation}
Although $O_0$ is not an eigenstate of the dilatation operator, its two-point function $\mathcal{G}^{<}_{00}
=-(\mathcal{G}^{<}_{+-}+\mathcal{G}^{<}_{-+})$ is real and given by:
\begin{align}
\label{eq:oscillateG}
\mathcal{G}^{<}_{00}(x)
=-|x^2|^{-\Delta_O-a}2  \Big[ 
&\mathrm{Re}(G^{<}_{+-})| \cos(\omega \log|x^2|) \nonumber\\
&\hspace{-1cm}-\mathrm{Im}(G^{<}_{+-}) \sin(\omega\log|x^2|)
\Big]\,,
\end{align}
This expression exhibits logarithmic oscillations, distinct from the monotonic power-law behavior of unitary correlators.
The oscillatory behavior is depicted in Figure~\ref{fig:oscill}.
In the language of the renormalization group, this implies that the RG flow of the corresponding operators exhibits limit-cycle behavior rather than conventional fixed points.

An even more intricate scaling behavior emerges when $N_c$ is \emph{precisely} at the exceptional point. 
At the EP, the two eigenvalues collide,
\begin{equation}
\lambda_{+}=\lambda_{-}\equiv \lambda_{\rm EP}\,,
\end{equation}
and the dilatation matrix assumes a non-diagonalizable Jordan block form.
The two eigen-operators ${\cal O}_{\pm}$ coalesce into a single state,
forcing the correlators to acquire a pure logarithmic scaling term, $\log|x^2|$.

We can derive this behavior by analyzing the limit as $N_c \rightarrow N_c^{\rm ep}$.%
Consider approaching $N_c^{\rm ep}$ from the $\mathcal{PT}$-broken phase, $N_c=N_{c}^{\mathrm{EP}}-\delta^2$ with $\delta\ll1$.
The eigenvalues and the non-vanishing Gram matrix elements have series expansions with respect to $\delta$ as
\begin{equation}
\label{eq:defineAM}
\lambda^{<}_{\pm}=a\pm\mathbbm{i}\omega \delta+{O}(\delta^2)\,,\quad
G^{<}_{+-}=-m\mathbbm{i}\delta-r\delta^2+{O}(\delta^3)\,, 
\end{equation}
where $a$, $\omega$, $m$, $r$ are real constants of ${O}(1)$.
Similarly, approaching $N_c^{\mathrm{ep}}$ from the $\mathcal{PT}$-unbroken phase ($N_c=N_{c}^{\mathrm{EP}}+\delta^2$), we have:
\begin{align}
&\lambda^{>}_{\pm}=a\pm \omega \delta+{O}(\delta^2)\,,\
&{G}^{>}_{\pm \pm}=\pm m \delta+r\delta^2+{O}(\delta^3)\,. \nonumber
\end{align}

The operator ${\cal O}_0$ defined in Eq.\,\eqref{eq:O0} vanishes linearly at the EP as ${O}(\delta)$.  
However, we can introduce a normalized operator:
\begin{equation}
\label{eq:tildeO}
{\tilde {\cal O}}_0:={\cal O}_0/\delta \,,
\end{equation}
which remains finite and is linearly independent of the coalesced eigen-operator in the limit $\delta\rightarrow0$.
Its two-point function in the two phases is given respectively by:
\begin{align}
{\mathcal{\tilde G}}_{00}^{<}&=-\frac{1}{\delta^2}(\mathcal{G}^<_{-+}+\mathcal{G}^<_{+-})\,,  \quad
\mathcal{\tilde G}^{>}_{00}=\frac{1}{\delta^2}(\mathcal{G}^>_{++}+\mathcal{G}^>_{--})\,.\nonumber
\end{align}
Taking the limit $\delta\rightarrow 0$, both expressions converge to the same \emph{finite} function:
\begin{equation}
\label{eq:logG}
\lim_{\delta \to 0} \mathcal{\tilde G}_{00}(x) =2|x^2|^{-\Delta_O-a} (r-\omega m\log|x^2|) \,.
\end{equation} 
This explicitly demonstrates that the logarithmic oscillation (broken phase) and the power-law scaling (unbroken phase) continuously merge into a pure logarithmic scaling violation at the EP.

The coalescence of operators at the EP is the precise mechanism by which the dilatation operator forms a rank-2 Jordan block, which is the defining algebraic signature of logarithmic conformal field theories (LCFTs) \cite{Gurarie:1993xq, Cardy:2013rqg}. 
At the EP, one can choose the operator basis consisting of the single eigen-operator (${\cal O}_{\rm ep}={\cal O}_{-}={\cal O}_{+}$) and $\tilde {\cal O}_0$.
On this basis, the dilatation matrix takes the canonical Jordan form. Concrete results for the dimension-8, length-4 sector are provided in Appendix~\ref{app:Jordan}.

\subsection{Non-abelian geometric phase}
\label{sec:geemetric}

The physics of exceptional points becomes particularly rich when $N_c$ is treated as a complex parameter. In the complex $N_c$ plane, the EPs appear as branch-point singularities of the dilatation matrix spectrum, giving the space of eigen-operators the structure of a multi-sheeted Riemann surface. This structure can be probed by transporting eigen-operators along closed paths in the complex $N_c$ plane. If a path encircles an EP, the eigenvectors undergo a nontrivial monodromy
\cite{Heiss:1999qe,Heiss:2012dx}: the final set of eigenvectors differs from the initial set by a linear transformation described by a monodromy matrix $\mathbb{M}$,
\begin{align}
\label{eq:defineM}
v^a\big|_{\mathrm{fin}}=\mathbb{M}_{ab}(\mathcal{C})\cdot v^b\big|_{\mathrm{ini}}\,,
\end{align}
which permutes the operator basis.

\begin{figure}
\centering
{\includegraphics[width=0.96\linewidth]{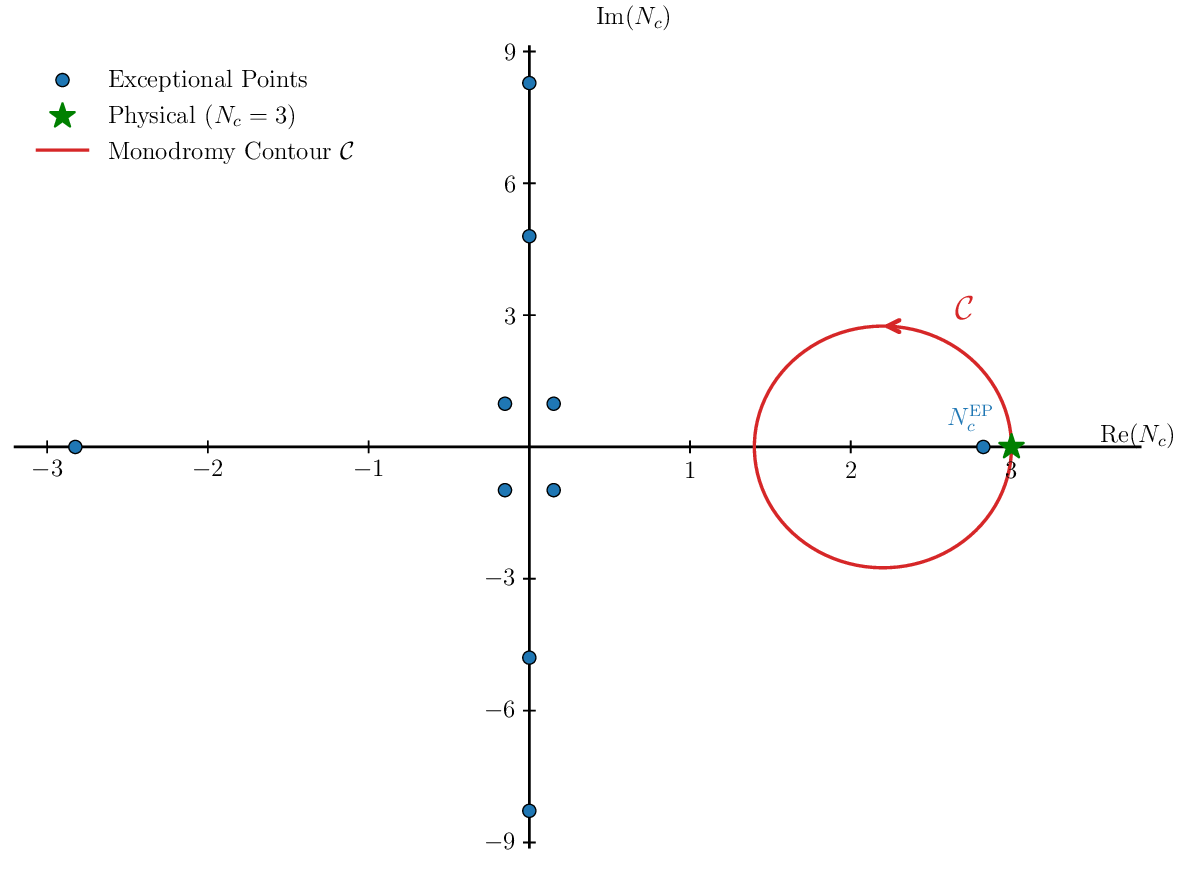}
}
\caption{\label{fig:contour} Blue points are exceptional points. The closed path $\mathcal{C}$ encircling
the $N_c^{\rm ep}=2.82466$. }
\end{figure} 

For the $(-)^4$ helicity sector of dimension-8 operators defined in Eq.\,\eqref{eq:newbaselen4}, 
consider a path $\mathcal{C}$ that encircles the single real EP at $N_c^{\rm EP}=2.82466$, as shown in Figure~\ref{fig:contour}. 
This path induces the monodromy matrix
\begin{align}
\label{eq:mono1}
\mathbb{M}(\mathcal{C})=\begin{pmatrix}
1 & 0 & 0 & 0\\
0 & 1 & 0 & 0\\
0 & 0 & 0 & -1\\
0 & 0 & 1 & 0
\end{pmatrix}\,.
\end{align}
This matrix permutes the two eigen-operators ${\cal O}_{+}$ and ${\cal O}_{-}$ that coalesce at the EP, up to a sign change.
This permutation reflects the multi-valued analytic structure of correlation functions under continuation in complex $N_c$:
analytic continuation around the EP exchanges the corresponding operator branches,
\begin{align}
\langle {\cal O}_{+} {\cal O}_i\cdots {\cal O}_j\rangle\rightarrow
\langle {\cal O}_{-} {\cal O}_i \cdots {\cal O}_j\rangle\,.
\end{align}
We find that the fourth power of the monodromy matrix is the identity, $\mathbb{M}^4=1$. This is consistent with the characteristic ``four-fold winding" structure of a standard second-order EP, a universal feature observed in diverse physical systems, see e.g.~\cite{Dembowski:2001zz, dembowski2004encircling}.

The full landscape of EPs in the complex $N_c$ plane for this system is shown in Figure~\ref{fig:contour}.  
The EPs appear as conjugate pairs, reflecting the fact that the discriminant in Eq.\,(\ref{eq:discrim1}) is a polynomial in $N_c^2$.
Encircling different EPs generates distinct permutations between subsets of eigen-operators. 
For a given operator sector, one can enumerate all the possible permutations
by exploring the group generated by all the EP-generated monodromy matrices. 
In the language of topological physics, this realizes a non-Abelian braiding of the operator spectrum \cite{Wang:2021nature}.
Consider the physical theory at $N_c=3$.
If we allow $N_c$ to vary along closed contours in the complex $N_c$-plane that 
start and end at $N_c=3$, the monodromy group acts on the set of physical operators. 
For the 4-operator sector discussed above, we find that by choosing appropriate contours that encircle specific EPs, one can realize the interchange of any two eigen-operators.
This implies that what appear to be distinct operators at $N_c=3$ should be viewed as different local branches of a single multi-valued analytic structure defined over the complex $N_c$ manifold.

This monodromy is a direct manifestation of a non-Abelian geometric phase. It can be understood from a differential-geometric perspective as the result of the parallel transport of eigenvectors in the parameter space.
The variation of the eigenvectors in the complex $N_c$ plane defines a non-Abelian Berry connection ${\cal A}_I$ in the parameter space (see e.g., \cite{Berry:2004ypy, Mailybaev:2005eet, vanderbilt2018berry, Kawabata:2018gjv, Ding:2022juv})
\begin{align}
\label{eq:omega2}
({\cal A}_I)^{ab}:= -( u^b)^{\mathrm{T}} \partial_I v^a \,,
\end{align}  
where $u^b$ and $v^a$ are the left and right eigenvectors of the 
dilatation matrix (see Appendix~\ref{app:geophase}), and $I=1,2$ indexes the coordinates of the complex $N_c$ space.
We parameterize the complex $N_c$-plane by real coordinates $\xi^1$, $\xi^2$.
The monodromy matrix for any path $\mathcal{C}$ is simply the path-ordered exponential of this connection: 
\begin{align}
\label{eq:wilsonLine}
\mathbb{M}=\mathcal{P}\exp(\oint_{\mathcal{C}} {\cal A}_I \,d\xi^I)\,.
\end{align}
In this language, the EPs appear as singularities of the Berry curvature in parameter space, 
and monodromy matrices correspond to Wilson loops of the non-Abelian connection around these singular points.

\section{Discussion} \label{sec:con}

In this work, we demonstrate that a non-Hermitian structure emerges upon analytic continuation of standard YM theory.
This result bridges two distinct frontiers: the rigorous renormalization program of high-energy gauge theories and the topological phenomenology of non-Hermitian physics.

The central mechanism driving this phenomenon is the interplay between the operator algebra and the complexified parameter space.
We identify \emph{color-evanescent operators}---operators that vanish at specific integer $N_c$ due to trace identities---as the origin of an indefinite metric in the operator space. 
When $N_c$ is treated as a continuous variable, these operators persist and introduce negative-norm directions in the operator inner-product space. 
Consequently, the dilatation operator $\operD$ (derived from the renormalization matrix), acting as the effective Hamiltonian, becomes non-Hermitian with respect to this inner product.

Our full-color analysis up to two-loop order reveals that this non-Hermiticity possesses a rich topological structure characterized by Exceptional Points (EPs). The emergence of these EPs has profound physical implications:
\begin{itemize}
\item 
\textbf{Spacetime Symmetry Breaking:}
We establish that the effective $\mathcal{PT}$ symmetry of the dilatation matrix is directly inherited from the fundamental spacetime $\mathcal{PT}$ symmetry of the gauge theory. The transition from real to complex anomalous dimensions at the EP corresponds to the spontaneous breaking of this symmetry.
\item 
\textbf{Logarithmic Scaling:} We identified logarithmic oscillations in two-point correlation functions, a hallmark of non-unitarity. Precisely at the EP, the dilatation matrix assumes a Jordan block form, leading to logarithmic scaling violations analogous to those found in logarithmic conformal field theories (LCFTs) \cite{Gurarie:1993xq, Cardy:2013rqg}. 
This indicates that analytic continuation of YM theory connects continuously to LCFT-like structures in complex parameter space, without modifying the underlying Lagrangian.
\item 
\textbf{Topological Monodromy:} We showed that EPs act as topological defects in the complex $N_c$ plane. Analytic continuation along a closed contour encircling an EP induces a non-Abelian geometric phase, resulting in a nontrivial monodromy  that permutes the identity of physical operators on different Riemann sheets of the spectrum. 

\end{itemize}

We stress that our study differs fundamentally from previous investigations of EPs in quantum field theories that begin with a non-Hermitian Lagrangian (see e.g.~\cite{Bender:2004sv, Kazakov:2022dbd, Bender:2021fxa}). 
In contrast, we consider the standard YM theory, which is ostensibly unitary. 
The non-Hermitian physics arises solely from the analytic continuation of a fundamental constant, the number of colors $N_c$, rather than from any deformation of the microscopic action.

\subsection*{On the interpretation of non-integer $N_c$}

A natural question arises regarding the physical interpretation of a gauge theory with non-integer rank.
We comment below on the utility and significance of such studies.

First, the mathematical consistency of this procedure is grounded in the framework of Deligne categories \cite{Deligne}, which generalize the concept of the Lie group. 
In this framework, color-evanescent operators correspond to objects associated with null idempotents---representations that vanish at integer $N_c$ but persist with indefinite norm for generic complex $N_c$, analogous to the study of \cite{Binder:2019zqc}. 
While \cite{Binder:2019zqc} provides the algebraic justification for non-unitarity, our work reveals the `dynamical' consequences in the operator spectrum: the emergence of EPs, the spontaneous breaking of $\mathcal{PT}$ symmetry, and the topological monodromy.

Second, physically, the analytical continuation of the rank of Lie groups has already played a significant role in statistical physics. 
As mentioned in the introduction, the O($N)$ vector model in the limit $N \rightarrow0$ describes self-avoiding polymers \cite{deGennes:1972zz}, while loop models generalize O($N$) spin models to arbitrary continuous $N$ \cite{Nienhuis:1982fx}. 
In these systems, non-integer $N$ corresponds to non-local loop fugacities.
By analogy, the non-integer $N_c$ in YM theory may be interpreted as generalizing discrete color degrees of freedom to continuous values with a similar loop or string-like interpretation, specifically, analytic continuation trades local degrees of freedom (colors) for non-local topological weights.

Third, the large $N_c$ expansion  \cite{'tHooft:1973jz} is itself a continuation of $N_c$ and has been critical to the study of gauge theories. 
The standard $1/N_c$ expansion implicitly assumes that the analytic continuation from $N_c=\infty$ to small integers such as $N_c=3$ is smooth.
Our results indicate that this analytic continuation can encounter branch-point singularities at finite complex $N_c$, and the radius of convergence in the expansion parameter $1/N_c$ is controlled by the distance to the nearest EP.

As discussed above, the dimension-8 length-4 operators induce EPs in the region $2<N_c<3$, so the $1/N_c$ series remains convergent for physical QCD ($N_c=3$) in this sector.
However, the inclusion operators of higher length and higher mass dimension can modify this structure.
For example, in the dimension-12, length-5 sector, we observe EPs in the region $3<N_c<4$, , which renders the standard planar expansion divergent for these operators at $N_c=3$.
More generally, as the operator complexity (length and mass dimension) increases, the associated EPs extend further outward in the complex $N_c$ plane. This reveals a fundamental limitation of the large-$N_c$ limit: its convergence is not uniform across the operator spectrum. A systematic study of this phenomenon would be of considerable interest.

Finally, one may expect that ``unphysical" singularities govern the analytic structure of the entire theory, thereby constraining the physics at integer $N_c$.
The geometric phase accumulated by circling EPs provides a concrete realization of this effect.

\subsection*{Outlook}
This study opens a new frontier at the intersection of quantum field theory and non-Hermitian physics. Several promising directions emerge:
\begin{itemize}

\item 
While we focused on the color charge in this work, this mechanism should apply to general QFTs. It would be illuminating to investigate the analytic continuation of other discrete parameters, such as flavor numbers (see e.g.~\cite{Maldacena:2011jn, Shimada:2015gda, Binder:2019zqc}), or multiple parameters simultaneously. 

\item
Our study is perturbative in the gauge coupling. However, since the geometric phase associated with EPs is topological, we expect these structures to persist non-perturbatively.
Verifying this in $\mathcal{N}=4$ Super Yang-Mills theory beyond the planar limit \cite{Beisert:2010jr} would be an important test, potentially linking EPs to integrability breaking.

\item
In quantum spectra, the distribution of energy levels characterizes the transition from integrability to chaos \cite{Bohigas:1983er}.
It would be interesting to explore whether the distribution of EPs in the complex $N_c$ plane for high-dimension operators would reveal hidden patterns or ``phases" of the gauge theory.

\item 
In non-Hermitian physics, the splitting of eigenvalues near an EP scales as $\Delta E \propto \sqrt{\Delta \lambda}$, offering enhanced sensitivity compared to the linear scaling of Hermitian systems \cite{2014PhRvL.112t3901W, Chen2017}.
It is intriguing to speculate whether this effect manifests in QFT, where physical observables near an EP (in parameter space) might show amplified sensitivity to `external' perturbations such as coupling variations.

\item
While all EPs identified in this paper are second order (where two operator eigenstates coalesce), the existence of higher-order EPs in fundamental QFTs remains an intriguing open question. Such higher-order structures have been extensively studied in non-Hermitian QM or optical systems (see e.g.~\cite{Heiss_2008, Graefe:2008shp, Hodaei:2017}) and exhibit even richer topological properties. A phenomenological application of third order EP in a holographic context was recently considered in \cite{Ghodrati:2025fah}.

\end{itemize}

\vskip .5cm
\textit{Acknowledgements.---} 
We would like to thank Bo Feng, Tao Shi, and Huajia Wang for discussions. 
This work is supported in part by the National Natural Science Foundation of China (Grants No.~12425504, 12447101, 12247103) and by the Chinese Academy of Sciences (Grant No. YSBR-101).
We also acknowledge the support of the HPC Cluster of ITP-CAS.

\appendix

\section{Details of length-4 operators}
\label{app:operators}

In this appendix, we provide the explicit definitions of the dimension-8, length-4 operators used throughout the main text. 
We then detail the classification of color factors and the construction of the operator basis for general operators.

\subsection{Length-4 dimension-8 operators}

The eight independent dimension-8 length-4 operators are classified into two helicity sectors: $(-)^4$ and $(-)^2(+)^2$.

The four basis operators in the $(-)^4$ sector are:
\begin{widetext}
\begin{equation}
\begin{aligned}
\label{eq:newbaselen4}
&\symbolOb_{1}=-\frac{1}{4}\delta^{i_1\cdots i_4}_{j_1\cdots j_4} 
(T^a_{i_1j_1}\cdots T^d_{i_4j_4})  (F^a_{12}F^b_{23}F^c_{34}F^d_{14}+
\frac{3}{4}F^a_{12}F^b_{12}F^c_{34}F^d_{34})
\,,
\\
&\symbolOb_{2}=\frac{1}{2}\delta^{i_1\cdots i_4}_{j_1\cdots j_4}(T^a_{i_1j_1}\cdots T^a_{i_4j_4})  (F^a_{12}F^b_{23}F^c_{34}F^d_{14}+\frac{1}{4}F^a_{12}F^b_{12}F^c_{34}F^d_{34}) 
-\delta^{i_1 i_2 i_3}_{j_1 j_3 j_4}\delta^{i_4}_{j_2}
(T^a_{i_1j_1}\cdots T^d_{i_4j_4})  F^a_{12}F^b_{23}F^c_{34}F^d_{14}
\,,
\\
&\symbolOb_{3}=\mathrm{tr}(F_{12}F_{23})\mathrm{tr}(F_{34}F_{14})
 +\frac{1}{2}\mathrm{tr}(F_{12}F_{34})\mathrm{tr}(F_{12}F_{34})
 +\frac{1}{4}[\mathrm{tr}(F^2)]^2\,,
\\
&\symbolOb_{4}=\mathrm{tr}(F_{12}F_{34})\mathrm{tr}(F_{23}F_{14})
+\frac{1}{2}\mathrm{tr}(F_{12}F_{34})\mathrm{tr}(F_{12}F_{34})
 +\frac{1}{4}[\mathrm{tr}(F^2)]^2 \,.
\end{aligned}
\end{equation}
\end{widetext}

The four basis operators in the $(-)^2(+)^2$ sector are:
\begin{widetext}
\begin{equation}
\begin{aligned}
\label{eq:newbaselen4b}
&\symbolOb_{5}=\frac{-1}{4}\delta^{i_1\cdots i_4}_{j_1\cdots j_4} 
(T^a_{i_1j_1}\cdots T^d_{i_4j_4})(F^a_{12}F^b_{23}F^c_{34}F^d_{14}+
\frac{1}{4}F^a_{12}F^b_{12}F^c_{34}F^d_{34})
\,,
\\
&\symbolOb_{6}=\frac{1}{2}\delta^{i_1\cdots i_4}_{j_1\cdots j_4}(T^a_{i_1j_1}\cdots T^a_{i_4j_4})
(F^a_{12}F^b_{23}F^c_{34}F^d_{14}+\frac{3}{4}F^a_{12}F^b_{12}F^c_{34}F^d_{34})
\\
&\qquad\ -\delta^{i_1 i_2 i_3}_{j_1 j_3 j_4}\delta^{i_4}_{j_2}
(T^a_{i_1j_1}\cdots T^d_{i_4j_4})  (F^a_{12}F^b_{23}F^c_{34}F^d_{14}
+F^a_{12}F^b_{12}F^c_{34}F^d_{34})\,,
\\
&\symbolOb_{7}=\mathrm{tr}(F_{12}F_{34})\mathrm{tr}(F_{23}F_{14})
+\frac{1}{2}\mathrm{tr}(F_{12}F_{34})\mathrm{tr}(F_{12}F_{34}) -\frac{1}{4} [\mathrm{tr}(F^2)]^2 \,,
\\
&\symbolOb_{8}=\mathrm{tr}(F_{12}F_{23})\mathrm{tr}(F_{34}F_{14})
+\frac{1}{4}[\mathrm{tr}(F^2)]^2  \,.
\end{aligned}
\end{equation}
\end{widetext}


\subsection{General classifications}
\label{app:classifyCF}
 
As introduced in Section~\ref{sec:setup}, Kronecker symbols in color space
can be contracted with Lie algebra generators to produce basis color factors, 
such as the ones in (\ref{eq:delta3color}).
We construct the complete basis of color factors by demanding that each length-$n$ color factor be a product of Kronecker symbols contracted with a tuple of generators:
\begin{align}
\label{eq:generalCF}
\delta^{i_{\sigma(1)}\cdots i_{\sigma(a)}}_{
j_{\tau(1)}\cdots j_{\tau(a)}}\cdots 
\delta^{i_{\sigma(b)}\cdots i_{\sigma(n)}}_{
j_{\tau(b)}\cdots j_{\tau(n)}}
\, T^{a_1}_{i_1j_1}\cdots T^{a_n}_{i_n j_n}\,,  
\end{align}
where $\sigma,\tau\in S_n$. If the highest-rank of Kronecker symbol in this product is $k$, we say this color 
factor is a $\delta_k$ color factor, which is mentioned in Section~\ref{sec:setup}. 
For example, $\delta^{i_1 i_2 i_3}_{j_1 j_2 j_4}\delta^{i_4}_{j_3}
(T^{a_1})_{i_1j_1}\cdots T^{a_4}_{i_4 j_4}$ is a $\delta_3$ factor.

Operators can be classified according to their charge conjugation $C$ parity. The $C$-parity operation $\hat{C}$ acts on the gauge field as:
\begin{align}
A_{\mu}^a T^a\xlongrightarrow{\hat{C}}
-A_{\mu}^a (T^a)^{\mathrm{T}}\,.
\end{align}
We define $C$-even (odd) operators as eigenstates of $\hat{C}$ with eigenvalues $+1(-1)$.
Operators with different $C$-parities never mix, so both the Gram and dilatation matrices are block-diagonal respecting with respect to this quantum number.

For length-4 operators, there are six single-trace basis elements $\mathrm{tr}(ijkl)$ and three double-trace elements $\mathrm{tr}(ij)\mathrm{tr}(kl)$.
Here $\mathrm{tr}(i\cdots j)$ is the abbreviation for $\mathrm{tr}(T^{a_i}\cdots T^{a_j})$.
The double-trace operators are all $C$-even.
The six single-trace operators can be grouped into three $C$-even ones and three $C$-odd combinations:
\begin{align}
& \mathrm{tr}(1\sigma(2)\sigma(3)\sigma(4))
+\mathrm{tr}(\sigma(4)\sigma(3)\sigma(2)1) \,, \\
& \mathrm{tr}(1\sigma(2)\sigma(3)\sigma(4))
-\mathrm{tr}(\sigma(4)\sigma(3)\sigma(2)1) \,,
\end{align}
where $\sigma$ represents a cyclic permutation in $Z_3$.

A useful observation for the general form in Eq.\,(\ref{eq:generalCF}) is that interchanging the upper and lower indices $(\sigma\leftrightarrow\tau)$ is equivalent to taking the transpose of all generators. 
Thus, one can construct the $C$-even and $C$-odd color factors by contracting
\begin{equation}
\begin{aligned}
\label{eq:generalCF2}
&\delta^{i_{\sigma(1)}\cdots i_{\sigma(a)}}_{
j_{\tau(1)}\cdots j_{\tau(a)}}\cdots 
\delta^{i_{\sigma(b)}\cdots i_{\sigma(n)}}_{
j_{\tau(b)}\cdots j_{\tau(n)}}
+(\sigma\leftrightarrow\tau)\,,
\\
&\delta^{i_{\sigma(1)}\cdots i_{\sigma(a)}}_{
j_{\tau(1)}\cdots j_{\tau(a)}}\cdots 
\delta^{i_{\sigma(b)}\cdots i_{\sigma(n)}}_{
j_{\tau(b)}\cdots j_{\tau(n)}}
-(\sigma\leftrightarrow\tau) \,,
\end{aligned}
\end{equation} 
with $T^{a_1}_{i_1j_1}\cdots T^{a_n}_{i_n j_n}$.
For example, when $n$ is even, the first symmetric combination is $C$-even and the second is $C$-odd.

We construct the basis hierarchically: first select the unique $\delta_L$ color factor (since all $\sigma, \tau$ permutations are equivalent), then consider all independent $\delta_{L-1}$ ones, and so on.
For length-4, the unique $\delta_4$ factor is
\begin{align}
\label{eq:L4D40}
&\delta^{i_1i_2i_3i_4}_{j_1 j_2 j_3 j_4}
T^{a_1}_{i_1 j_1} T^{a_2}_{i_2 j_2} T^{a_3}_{i_3 j_3}T^{a_4}_{i_4 j_4}
\\
&=-\sum_{\sigma\in S_3}\mathrm{tr}(1\sigma(2)\sigma(3)\sigma(4))
+\sum_{\tau\in Z_3}\mathrm{tr}(1\tau(2))\mathrm{tr}(\tau(3)\tau(4))\,, \nonumber
\end{align}
which is $C$-even.

The linearly independent $\delta_3$ factors factors include two $C$-even combinations:
\begin{align}
\label{eq:L4D301}
\big(\{
\delta^{i_1i_2i_3}_{j_1 j_2 j_4}\delta^{i_4}_{j_3},
\delta^{i_1i_2i_3}_{j_1j_3j_4}\delta^{i_4}_{j_2}\}+(i\leftrightarrow j)\big)
\times T^{a_1}_{i_1 j_1} T^{a_2}_{i_2 j_2} T^{a_3}_{i_3 j_3}T^{a_4}_{i_4 j_4}\,,
\end{align}
and three $C$-odd combinations:
\begin{align}
\label{eq:L4D302}
&\big(\{
\delta^{i_1i_2i_3}_{j_1 j_2 j_4}\delta^{i_4}_{j_3},
\delta^{i_1i_2i_3}_{j_2j_3j_4}\delta^{i_4}_{j_1} ,
\delta^{i_1i_2i_4}_{j_2j_3j_4}\delta^{i_3}_{j_1} 
\}-(i\leftrightarrow j)\big)
\nonumber\\
&\times T^{a_1}_{i_1 j_1} T^{a_2}_{i_2 j_2} T^{a_3}_{i_3 j_3}T^{a_4}_{i_4 j_4}\,.
\end{align}
Their relation with trace bases can be seen by expand the first terms of Eq.\,(\ref{eq:L4D301}) and Eq.\,(\ref{eq:L4D302}):
\begin{align}
&(\mathrm{tr}(1234)+\mathrm{tr}(1243)-\mathrm{tr}(12)\mathrm{tr}(34))+\mbox{color reversing}\,,
\nonumber\\
&(\mathrm{tr}(1234)+\mathrm{tr}(1342))-\mbox{color reversing}\,.
\nonumber
\end{align}
%
The remaining three independent $\delta_2$ factors correspond to the three double-trace structures, which are all $C$-even:
\begin{align}
\label{eq:L4D20}
&\delta^{i_1 i_{\tau(2)}}_{j_1  j_{\tau(2)}} 
\delta^{ i_{\tau(3)} i_{\tau(4)}}_{ j_{\tau(3)} j_{\tau(4)}}
\times
T^{a_1}_{i_1 j_1} T^{a_2}_{i_2 j_2} T^{a_3}_{i_3 j_3}T^{a_4}_{i_4 j_4} \\
& =
\mathrm{tr}(1\tau(2))\mathrm{tr}(\tau(3)\tau(4))
\,,
\quad \tau\in Z_3\,. \nonumber
\end{align}
For the dimension-8 operators, contracting these color factors with the kinematic invariants  
\begin{equation}
\begin{aligned}
&(F^{a}_{12} F^{b}_{23} F^{c}_{34} F^{d}_{14})\,,\quad
(F^{a}_{12}F^{b}_{12}F^{c}_{34} F^{d}_{34})\,,
\end{aligned}
\end{equation}
we obtain one $\delta_4$, one $\delta_3$ and two $\delta_2$ operators for each kinematic factor, respectively.
In this special case, $C$-odd operators vanish due to kinematic symmetry, leaving 8 $C$-even operators.

To get the basis presented in Eq.\,\eqref{eq:newbaselen4} and Eq.\,\eqref{eq:newbaselen4b}, the 8 operators are further reorganized into two helicity sectors: $(-)^4$ and $(-)^2(+)^2$. 
Here, an operator belongs to a certain helicity sector $(-)^\alpha(+)^\beta$ means that its tree-level minimal form factors at $d=4$ is nonzero if and only if the external gluons take the corresponding helicity configuration (or the conjugate one
$(+)^\alpha(-)^\beta$).  

We also analyze length-5 operators up to dimension 12. The explicit operator bases, along with their corresponding Gram and one-loop dilatation matrices, are provided in the ancillary files. The counting of independent length-5 color factors is summarized by the following table:
\begin{equation}
\begin{tabular}{|c|c|c|c|c|   }
\hline
  & $\delta_5$ & $\delta_4$ & $\delta_3$ & $\delta_2$ \\
\hline
$C$-even & 0  & 6  & 10 & 6 \\
\hline
$C$-odd & 1 &  5 & 16 & 0 \\
\hline
\end{tabular}
\end{equation}
The unique $\delta_5$ factor is $C$-odd. Since a $\delta_n$ operator is color-evanescent for any integer $N_c<n$, we expect EPs  to emerge in the range $3<N_c<4$ to appear specifically within $C$-odd length-5 sectors. 
Our explicit one-loop calculations confirm this expectation.

Finally, we note that operators can be further classified by their Lorentz derivative structures.
We define an operator to be of type $D$-$(i,\alpha)$ if it can be written as the $i$th 
derivative of a rank-$\alpha$ tensor operator. 
For example, $\partial_\mu \partial_\nu \mathrm{tr}({\cal O}^{\mu\nu})$ 
is of type $D$-(2,2) while $\partial^2\mathrm{tr}({\cal O})$ is 
of type $D$-(2,0).
The bases defined in Eq.\,(\ref{eq:newbaselen4}) and Eq.\,(\ref{eq:newbaselen4b}) are both of type $D$-(0,0).
The 16-element dimension-12 sector discussed in Section~\ref{sec:TwoEvaTypes} is of type $D$-(2,2).
We establish a hierarchy among these types: $D$-$(i,\alpha) > D$-$(j,\beta)$ if $i>j$, or
if $i=j$, $\alpha<\beta$.
Under renormalization, higher $D$-type operators do not mix into lower ones.
Therefore, organizing the operator basis according to this $D$-type hierarchy ensures that the dilatation matrix takes a manifest block-triangular form.

\section{IR divergences for form factors}
\label{app:IR}

In this appendix, we detail the structure of infrared (IR) divergences used to extract the ultraviolet renormalization constants in Section~\ref{subsec:dilatation}.

The one-loop IR divergences take the following structure  \cite{Catani:1998bh}:
\begin{align}
\label{eq:catani}
&\mathbf{F}^{(1)}_{\mathcal{O},\mathrm{IR}}  =
\mathbf{I}_{\rm IR}^{(1)} (\epsilon)
\mathbf{F}^{(0)}_{\mathcal{O}} , \\
&
\mathbf{I}_{\rm IR}^{(1)}(\epsilon)= \frac{e^{\gamma_E\epsilon}}{\Gamma(1-\epsilon)}
\big(\frac{1}{\epsilon^2}+\frac{\beta_0}{2 C_A\epsilon}\big)
\sum_{i<j}(-s_{ij})^{-\epsilon} \mathbf{T}_i\cdot \mathbf{T}_j\,, \nonumber
\end{align}
where $\mathbf{T}^a_i$ is color generator act on the $i$th gluon. The color dipole operator $\mathbf{T}_i\cdot \mathbf{T}_j$ acts on the color factor as:
\begin{align}
&\mathbf{T}_i\cdot \mathbf{T}_j\ \mathrm{tr}(\cdots T^{a_i} \cdots T^{a_j} \cdots) \\
&=\sum_b\mathrm{tr}(\cdots [T^b,T^{a_i}] \cdots [T^b,T^{a_j}] \cdots)\,.\nonumber
\end{align}

At two loops, the IR divergence structure is given by \cite{Catani:1998bh, Aybat:2006mz}:
\begin{align}
\mathbf{F}^{(2)}_{\mathcal{O},\mathrm{IR}}  =
\mathbf{I}_{\rm IR}^{(2)} (\epsilon)
\mathbf{F}^{(0)}_{\mathcal{O}} +\left(\mathbf{I}_{\rm IR}^{(1)} (\epsilon)
\mathbf{F}^{(1)}_{\mathcal{O},R}\right)\bigg|_{\mathrm{div}}\,, 
\label{eq:catani2loop}
\end{align}
where $\mathbf{F}^{(1)}_{\mathcal{O},R}$ is the finite, renormalized one-loop form factor, and $\mathbf{I}_{\rm IR}^{(1)} (\epsilon)$ is given in Eq.\,\eqref{eq:catani}. The two-loop operator $\mathbf{I}_{\rm IR}^{(2)} (\epsilon)$ is defined as:
\begin{align}
\mathbf{I}_{\rm IR}^{(2)} (\epsilon)=\ & -\frac{1}{2}(\mathbf{I}_{\rm IR}^{(1)} (\epsilon))^2-\frac{\beta_0}{\epsilon}\mathbf{I}_{\rm IR}^{(1)} (\epsilon) \\
& +\frac{e^{-\gamma_E\epsilon}\Gamma(1-2\epsilon)}{\Gamma(1-\epsilon)}(\frac{\beta_0}{\epsilon}+\mathcal{K})\mathbf{I}_{\rm IR}^{(1)} (2\epsilon) \nonumber\\
& +n \frac{e^{\gamma_e}}{\epsilon\Gamma(1-\epsilon)}\mathcal{H}^{(2)}_{\Omega,g}+\mathbf{H}^{(2)}\,, \nonumber
\end{align}
where
\begin{align}
	&\mathcal{K}=(\frac{67}{9}-\frac{\pi^2}{3})N_c, \quad \mathcal{H}^{(2)}_{\Omega,g}=(\frac{\zeta_3}{2}+\frac{5}{12}+\frac{11\pi^2}{144})N_c^2\,,\\
	&\mathbf{H}^{(2)}=\mathbbm{i}\sum_{1\leq i<j<k\leq n}f^{abc}\mathbf{T}^{a}_i\mathbf{T}^{b}_j\mathbf{T}^{c}_k\log{\frac{s_{ij}}{s_{kj}}}\log{\frac{s_{jk}}{s_{ki}}}\log{\frac{s_{ki}}{s_{ij}}}\,.
\end{align}

\section{Symmetrizing Dilatation}
\label{app:symH}

In this appendix, we provide the mathematical details for constructing the symmetrized dilatation matrix in Eq.\,(\ref{eq:symH}).

The relation between the Gram matrix $G$ and the dilatation matrix $\mathbb{D}$ follows from the hermitianity of the dilatation operation $\hat{D}$ acting on the Hilbert space:
\begin{equation}
\langle O_i |(\hat{D} O_j) \rangle=\langle (\hat{D}O_i) | O_j  \rangle\,.
\end{equation}
In terms of the matrix representations defined by $\hat{D}O_j = \sum_k \mathbb{D}_j^{\ k} O_k$, and $G_{ij}=\langle O_i | O_j\rangle$,
this condition becomes:
\begin{equation}
\label{eq:GD-DG}
G_{ik}\mathbb{D}_j^{\ k} =
\mathbb{D}^{*k}_{i} G_{kj}\,.
\end{equation}
In our basis, the coefficients of all monomial operators are real, ensuring that $G$ is real symmetric and $\mathbb{D}$ is real. Consequently, the matrix product 
$\mathbb{D}\cdot G$ is symmetric.

It is instructive to clarify the nature of our YM system compared to standard quantum mechanics. In standard QM, one works in a fixed Hilbert space equipped with a positive-definite inner product (where the Gram matrix is effectively the identity), and the Hamiltonian is Hermitian with respect to that inner product. In contrast, in our setting the inner product is defined by two-point correlation functions of composite operators, leading to a nontrivial $N_c$-dependent metric $G$. The dilatation matrix $\operD$ satisfies the generalized Hermiticity condition as given in Eq.\,\eqref{eq:GD-DG}. This identifies the system $(\operD,G)$
 as a \emph{pseudo-Hermitian} operator system \cite{Mostafazadeh:2001jk, Mostafazadeh:2008pw}, in which the metric $G$ is not an externally imposed structure but is dynamically determined by the gauge theory itself.

One can follow the standard method, \emph{e.g.}~Gram-Schmidt algorithm  \cite{cheney2009linear} 
to obtain a real matrix $\mathcal{R}$ such that:
\begin{align}
\label{eq:GramSchimdt}
\mathcal{R}\cdot G \cdot \mathcal{R}^{\mathrm{T}}=\mathcal{P}\,,
\end{align}
where $\mathcal{P}$ is a diagonal matrix with entries $\pm$, representing the signature of $G$.  
We then define the transformation matrix
\begin{align}
\label{eq:orthogonal}
M:=\sqrt{\mathcal{P}}\cdot \mathcal{R}\,.
\end{align}
Note that $M$ may be complex matrix if $\mathcal{P}$ contains negative entries, and $M^*=\mathcal{P}\cdot M$.
By construction, $M$ satisfies:
\begin{align}
\label{eq:Mproperty}
M\cdot G\cdot M^{\mathrm{T}}=\mathbbm{1}\,,\quad
M^*\cdot G\cdot M^{\mathrm{T}}=\mathcal{P}\,.
\end{align}
Using this transformation, we define the symmetrized dilatation matrix: 
\begin{align}
\label{eq:symH2}
\mathbb{H} :=M\cdot \mathbb{D} \cdot M^{-1}
=M\cdot \mathbb{D} \cdot G  \cdot M^{\mathrm{T}}\,.
\end{align} 
This matrix possesses two key properties.
First, $\mathbb{H} $ is similar to $\mathbb{D} $ and therefore  
shares the same eigenvalues.
Second, $\mathbb{H} $ is symmetric since it is congruent to the symmetric matrix 
$\mathbb{D} \cdot G$ via the transformation $M$.

As a concrete example, for the length-4 dimension-8 operator sector in Eq.\,(\ref{eq:newbaselen4}), one can obtain one-loop $\mathbb{H}$ from Eq.\,(\ref{eq:Galpha}) and Eq.\,(\ref{eq:oldDila1}):
\begin{align}
\label{eq:H1alpha2}
& \mathbb{H}^{(1)}  = 4 N_c\times
\nonumber\\
& \scalebox{0.8}{
$\displaystyle
\begin{pmatrix}
 -1 & \frac{2\sqrt{10 b (3n+2)}}{-3 \sqrt{u}} & 0 & 0 \\
 * & -\frac{3 n+10}{6 (3n+2)} & \frac{\sqrt{3 a(n+2)} (2 v-u)}{(3n+2)\sqrt{2u v} } & 
   \frac{ \sqrt{15 a(n+2)(n+3)}  }{\sqrt{v (3n+2)} } \\
 * & * & \frac{10 n^2 u+24 n v-103 u+39 v+240}{3 n v(3n+2)} 
  & \frac{ \sqrt{5\,u(n+3)}   (2 a+u+v)}{-n v\sqrt{2(3n+2)} } \\
 * & * & * & \frac{7 n u+2 n v+12 u+4 v+80}{-3 n v} \\
\end{pmatrix}
$},
\end{align}
where $n=N_c$, and the factors $a$, $b$, $c$, $u$ are defined in Eq.\,(\ref{eq:notation1}), with
$v=11 N_c^2+39 N_c+22$. 

For $N_c>3$, the Gram matrix $G^{(0)}$ is positive definite, consequently, $\mathbb{H}^{(1)}$ is real and symmetric (Hermitian), guaranteeing a real spectrum.
The eigenvectors corresponding to distinct eigenvalues are orthogonal ($v_i^\dag v_j=0$).
For $N_c<3$, the indefinite signature of $G^{(0)}$ introduces imaginary factors into $M$, making $\mathbb{H}^{(1)}$ 
complex symmetric (anti-Hermitian). This leads to the non-orthogonality of eigenvectors. 
However, the non-Hermitianity is not sufficient to generate complex eigenvalues.
As seen in Figure~\ref{fig:d8L4sec}, there exists an interval $(N_c^{\rm ep},3)$,
where the spectrum remains strictly real despite the non-Hermiticity of the Hamiltonian.

\section{More on $\mathcal{PT}$ symmetry}
\label{app:PTmat}

In  Section~\ref{sec:PT}, we identified two representations of the dilatation matrices: the original matrix $\operD$, and its symmetrized counterpart, $\operH$.
Correspondingly, we defined two realizations of the $\mathcal{PT}$ symmetry: $\tilde{P}_{\mathrm{eff}}\tilde{T}_{\mathrm{eff}}$ acting on the original basis, and ${P}_{\mathrm{eff}}{T}_{\mathrm{eff}}$ acting on the symmetrized basis. The dilatation matrix $\operD$ is more directly related to the YM operator.
In this appendix, we provide details of how the effective $\mathcal{PT}$ symmetries are direct manifestations of the fundamental spacetime symmetry of the YM fields.

Consider an eigen-operator $\mathcal{O}(x)$ expanded in the basis $\{\mathcal{O}_i(x)\}$:
\begin{equation}
\mathcal{O}=v_i \mathcal{O}_i \,.
\end{equation}
The coefficients $v_i$ form a left eigenvector of $\mathbb{D}$, or equivalently,
a right eigenvector of $\mathbb{D}^{\mathrm{T}}$.
In the effective quantum system described by $\mathbb{D}^{\mathrm{T}}$ , the composite operator $\tilde{P}_{\mathrm{eff}} \tilde{T}_{\mathrm{eff}}$ acts on the state vector $v$ as complex conjugation (see Eq.~(\ref{eq:PToverCoe})):
\begin{align}
\label{eq:combinePT-3}
\tilde{P}_{\mathrm{eff}}\tilde{T}_{\mathrm{eff}}\circ( v_i \mathcal{O}_i )= 
(\tilde{P}_{\mathrm{eff}}\tilde{T}_{\mathrm{eff}}\circ v_i ) \mathcal{O}_i = v_i^* \mathcal{O}_i\,.
\end{align}

We now compare this to the action of the spacetime $\mathcal{PT}$ operator, denoted by 
$\hat{P}$ and $\hat{T}$.
The transformation properties of the elementary gauge field and derivative are summarized in Table~\ref{tab:PT0}.
Specifically, the partial derivative $\partial_\mu$ is $\hat{P}\hat{T}$ odd, while the gauge field
$A_\mu$ is $\hat{P}\hat{T}$ even.
Consequently,  both $D_\mu=\partial_\mu-\mathbbm{i}g A_\mu$ and 
$F_{\mu\nu}=\partial_\mu A_\nu-\partial_\nu A_\mu-\mathbbm{i}g[A_\mu,A_\nu]$
are $\hat{P}\hat{T}$ odd.

We now determine the parity of the composite YM operators.
First, the covariant derivatives do not affect the composite $\hat{P}\hat{T}$ parity of a Lorentz scalar operator, since they always appear in even numbers.
Second, a length-$L$ operator contains $L$ field strengths  $F_{\mu\nu}$, contributing a factor $(-1)^L$.
Third, the commutator relation 
\begin{align}
[D_\mu,D_\nu]\diamond =-\mathbbm{i}g [F_{\mu\nu},\diamond]\,,
\end{align}
implies that to maintain a consistent phase under $\mathcal{PT}$, we should associate a factor of $\mathbbm{i}$
with each field strength in the definition of the basis. Thus a length-$L$ basis operator has a factor $\mathbbm{i}^L$, which also 
contributes a $(-1)^L$ for the $\hat{P}\hat{T}$ parity.
Combining these effects, the total parity factor is $(-1)^{2L}=1$.
Thus, the YM basis operators we consider are $\hat{P}\hat{T}$ even:
\begin{equation}
\hat{P}\hat{T}\circ \mathcal{O}_i(t,x) 
= \mathcal{O}_i(-t,-x)\,.
\end{equation}

Applying this to the composite operator expansion:
\begin{align}
\label{eq:combinePT-4}
\hat{P}\hat{T}\circ\big( v_i \mathcal{O}_i(x^\mu)\big)
=v_i^*\big(\hat{P}\hat{T}\circ\mathcal{O}_i(x^\mu)\big)
= v_i^* \mathcal{O}_i(-x^\mu)\,.
\end{align}
Comparing this result with Eq.\,\eqref{eq:combinePT-3}, we see that the action of the composite $\hat{P}\hat{T}$ on Yang-Mills operators is equivalent to the action of $\tilde{P}_{\mathrm{eff}}\tilde{T}_{\mathrm{eff}}$ on the state vectors in the effective quantum system (modulo the coordinate reflection).
This demonstrates that the emergent $\mathcal{PT}$ symmetry of the dilatation matrix is a direct representation of the fundamental spacetime $\mathcal{PT}$ symmetry of the Yang–Mills theory.

\section{Construction of the $\mathcal{C}$ operator}
\label{app:Cmat}

In this appendix, we detail the construction of the $\mathcal{C}$ operator, which allows us to define a positive-definite $\mathcal{CPT}$ inner product in the $\mathcal{PT}$-unbroken phase \cite{Bender:2002vv}.

Consider a non-Hermitian Hamiltonian $\mathcal{H}$ with a discrete spectrum associated with a complete biorthonormal eigenbasis $\{ |\psi_i \rangle, |\phi_i\rangle\}$:
\begin{equation}
\label{eq:defineUV}
\mathcal{H} | \psi_i \rangle=E_i |\psi_i \rangle\,,\quad
 \mathcal{H}^\dag | \phi_i \rangle  =E_i^* | \phi_i \rangle\,,
\end{equation}
satisfying the orthonormality condition:
\begin{equation}
\label{eq:uv-norm}
\langle \phi_i | \psi_j \rangle=\delta_{ij}\,.
\end{equation}
If $\mathcal{H}$ is symmetric (as is the case for our symmetrized dilatation matrix $\operH$), we have: 
$|\phi_i \rangle=(|\psi_i \rangle)^*$.

We define the $\hat\eta$-matrix as \cite{Mostafazadeh:2001jk}:
\begin{equation}
\hat{\eta} = \sum_{n \in \mathbb{R}} |\phi_n \rangle \langle \phi_n |+ 
\sum_{n_+ \in \mathbb{C}} \Big( |\phi_{n_+}\rangle\langle \phi_{n_-}|+ 
|\phi_{n_-}\rangle\langle\phi_{n_+}|\Big) ,
\end{equation}
where the first sum runs over eigenstates with real eigenvalues, and the second sum runs over pairs of states $(\phi_{n_+}, \phi_{n_-})$ with complex-conjugate eigenvalues.
The inverse operator $\hat{\eta}^{-1}$ is similarly defined using the eigenvectors $|\psi_i\rangle$.

By construction, $\hat{\eta}$ is a Hermitian operator that satisfies the intertwining relation:
\begin{align}
\label{eq:pseudo-hermitian}
\hat{\eta}\cdot\mathcal{H}=\mathcal{H}^\dag \cdot \hat{\eta}\,.
\end{align}

In the $\mathcal{PT}$-broken phase, there exist complex eigen-energies.
Since $E_i \neq E_i^*$, the corresponding state must have zero norm with respect to the metric $\hat\eta$:
\begin{equation}
0=
\langle \psi_i |\hat{\eta} \mathcal{H}| \psi_i \rangle-
\langle \psi_i |\mathcal{H}^\dag \hat{\eta}| \psi_i \rangle
=(E_i-E_i^*)\langle \psi_i | \hat{\eta}| \psi_i\rangle\,.
\end{equation}
The existence of these zero-norm states signifies that  unitarity is 
intrinsically broken in the $\mathcal{PT}$-broken phase. 
We comment that at exact the EP, the coalesced states (with degenerate real eigenvalues) also have zero norm with respect to the metric $\hat\eta$.

In the $\mathcal{PT}$-unbroken phase, however, all the eigenvalues are real. Thus, the metric simplifies to:
\begin{equation}
\hat{\eta} = \sum_{n} |\phi_{n}\rangle \langle \phi_{n}|\,.
\end{equation}
In this regime, $\hat{\eta}$ is positive-definite. The inner product defined by $\hat\eta$ renders the basis $\{\psi_i\}$ orthonormal:
\begin{align}
\label{eq:vv-norm}
\langle \psi_i |\hat{\eta}| \psi_j \rangle
=\sum_k  \langle \psi_i | \phi_k\rangle\langle \phi_k| \psi_j \rangle=\delta_{ij}\,.
\end{align}
By introducing an effective charge conjugation operator:
\begin{align}
\mathcal{C}:=\mathcal{P}\hat{\eta}\,,
\end{align}  
we can define a new inner product, 
the $\mathcal{CPT}$-norm, which is positive definite in the $\mathcal{PT}$-unbroken phase:
\begin{equation}
\big(\mathcal{CPT} \psi_i\big) \cdot \psi_j=
\langle \psi_i| \mathcal{PC} |\psi_j \rangle=\langle \psi_i|\hat{\eta} |\psi_j \rangle = \delta_{ij}\,.
\end{equation}
For a symmetric Hamiltonian $\mathcal{H}$, the matrix representation of $\mathcal{C}$ is symmetric. 
The $\mathcal{C}$ operator satisfies the standard algebraic relations:
\begin{align}
[\mathcal{C},\mathcal{PT}]=0\,,\quad
\mathcal{C}^2=1\,,\quad
[\mathcal{C},\mathcal{H}]=0\,.
\end{align}

We now apply this formalism to the YM operator basis. Recall the orthonormal basis $\mathcal{O}^{\mathrm{p}}_i$
defined in Eq.\,(\ref{eq:defineOp1}), in which the Gram matrix is diagonal with entries $\mathcal{P}_{ij}$. 
At the level of the operator algebra, the above $\mathcal{CPT}$ adjoint corresponds to defining a $\mathcal{C}$-conjugated operator basis:
\begin{align}
\mathcal{O}^{\mathrm{cp}}_i:= \sum_j \mathcal{C}_{ji} \mathcal{O}^{\mathrm{p}}_j\,.
\end{align}
The new Gram matrix $G^{\mathrm{c}}_{ij}$ is obtained by computing the two-point function between the standard basis and the $\mathcal{C}$-conjugated basis:
\begin{align}
\langle \mathcal{O}^{\mathrm{p}\dag}_i(x) \mathcal{O}^{\mathrm{cp}}_j(0)\rangle
&= \sum_k \langle \mathcal{O}^{\mathrm{p}\dag}_i(x) \mathcal{O}^{\mathrm{p}}_k(0)\rangle \mathcal{C}_{kj}  
\nonumber\\
&=\frac{(\mathcal{P}\mathcal{C})_{ij}}{|x^2|^{\Delta_{\mathcal{O}_i}}}
=:\frac{G^{\mathrm{c}}_{ij}}{|x^2|^{\Delta_{\mathcal{O}_i}}}\,.
\end{align}
Since $\mathcal{P}\mathcal{C}=\hat{\eta}$ is positive definite by construction, the new metric $G^{\mathrm{c}}$ is positive definite.

As an explicit example, consider the dimension-8, length-4 sector (Eq.~(\ref{eq:newbaselen4})) at a point $N_c=3-\smallN^2 \ (\smallN\ll1)$, which lies in the $\mathcal{PT}$-unbroken phase but close to the transition. In this region, the parity matrix is $\mathcal{P} = {\rm diag}(-1,1,1,1)$.
The $\mathcal{C}$ matrix is obtained as:
\begin{equation}
\mathcal{C}=
\scalebox{0.78}{
$\displaystyle
\begin{pmatrix}
 -\frac{20410\smallN^2}{7803}-1 & -\frac{32}{17} \mathbbm{i} \sqrt{\frac{15}{11}}  \smallN   
 & -\frac{190}{51} \mathbbm{i} \sqrt{\frac{2}{3927}} \smallN   &
   -\frac{30}{17} \mathbbm{i} \sqrt{\frac{15}{119}} \smallN   \\
* & 
 \frac{7680  \smallN^2 }{3179}+1 & \frac{3040 \sqrt{\frac{10}{119}} \smallN^2}{9537} &
   \frac{7200  \smallN^2}{289 \sqrt{1309}} \\
* & * & \frac{36100 \smallN^2}{10214127}+1 
 & \frac{950 \sqrt{\frac{10}{11}} \smallN^2}{34391} \\
* & * & * & 
 \frac{6750 \smallN^2}{34391}+1 \\
\end{pmatrix}
$}
+{O}(\smallN^3)\,.
\end{equation}
The lower triangle is omitted due to symmetry. The resulting metric $\mathcal{PC}$ is Hermitian and positive definite, confirming the consistency of the construction.

While we have shown the equivalence between the effective $\mathcal{PT}$-symmetry and the fundamental spacetime $\mathcal{PT}$-symmetry, we comment that the ${\cal C}$ operator, which depends on the operator sectors, appears to have no direct correspondence to the spacetime charge conjugation symmetry.

\section{Logarithmic scaling near EP}
\label{app:Jordan}

We provide a concrete example to illustrate how the Jordan form leads to the logarithmic scaling of the two-point function discussed in Section~\ref{sec:log}.

Consider the perturbation expansion in $\delta$ for the dimension-8 length-4 $(-)^4$ sector at 
$N_c=N_c^{\mathrm{EP}}\pm\delta^2$ with $N_c^{\mathrm{EP}}=2.825$.
The Gram and dilatation matrices for this system 
are given by Eq.\,(\ref{eq:Galpha}) and Eq.\,(\ref{eq:oldDila1}).
The eigenvectors of  the dilatation matrix are denoted by $v_i^{\pm}$ ($i=1,2,3,4$).
As $\delta\rightarrow 0$, the vectors coalesce such that
$v_{3}^{\pm}-v_{4}^{\pm}\sim {O}(\delta)$.
In the basis of eigenvectors, one can read the values of $a$, $\omega$, $m$ and $r$
(defined in Eq.\,(\ref{eq:defineAM})) directly from the eigenvalues of $\mathbb{D}$
and the entries of $(v_i^*)^{\mathrm{T}}G v_j$. These are given as:
\begin{align}
&a=-6.85206 \,\alpha\,,\quad  \omega=-11.1310\, \alpha\,,\nonumber\\ 
&m=-5.73437 \kappa\,,\quad  r=-13.0021 \kappa\,.\nonumber
\end{align}
Here $\alpha=\frac{\alpha_s}{4\pi}$ and $\kappa=\frac{6144(n^2-1)}{n}\times 10^4$ with $n=N_c$.
Notice that $\kappa/10^4$ is just the overall factor appearing in Eq.\,(\ref{eq:Galpha}).
Plugging these parameters into the limit expression for $\tilde{\mathcal{G}}_{00}$ in Eq.\,(\ref{eq:logG}), one obtains
\begin{equation}\label{eq:G00}
\tilde{\mathcal{G}}_{00}=\kappa |x|^{13.704\alpha-2\Delta}(-26.004 -255.316\alpha \log|x| )\,.
\end{equation}

Alternatively, one can derive the above result by directly solving the renormalization group (RG)
equation in the new basis involving $\tilde{\mathcal{O}}_0$.
As an example, in the $\mathcal{PT}$-broken phase, the new basis is chosen as
\begin{equation}
\{v_1^{-},v_2^{-},v_3^{-},-\mathbbm{i}(v_3^{-}-v_4^{-})/\delta\} \,.
\end{equation}
Transforming the Gram and dilatation matrices into this basis and expanding them up to linear order in $\delta$, we have
\begin{footnotesize}
\begin{align}
G^{-}&=\kappa 
\scalebox{0.9}{
$\displaystyle
\begin{pmatrix}
 0.0976 & 0 & 0 & 0 \\
 0 & 0.231 & 0 & 0 \\
 0 & 0 & 0 & -5.734+13.002 i \delta  \\
 0 & 0 & -5.734-13.002 i \delta  & -26.004 \\
\end{pmatrix}
$}
+{O}(\delta^2),
\\
\mathbb{D}^{-}&=\alpha
\scalebox{0.9}{
$\displaystyle
\begin{pmatrix}
 -51.42 & 0 & 0 & 0 \\
 0 & 40.64 & 0 & 0 \\
 0 & 0 & -6.85 -11.13 i \delta  & 0 \\
 0 & 0 & -22.26 & -6.85+11.13 i \delta  \\
\end{pmatrix}
$}
+{O}(\delta^2)\,\,,
\end{align}
\end{footnotesize}
where the dilatation matrix takes a Jordan form in the limit $\delta\rightarrow0$.

Solving the RG equation 
\begin{align}
\frac{d\mathcal{G}_{ij} }{d\log|x|}+2\Delta \mathcal{G}_{ij}
+ (\mathbb{D}_i^{\,k})^*\mathcal{G}_{kj} 
+ \mathbb{D}_j^{\,k}\mathcal{G}_{ik}=0\,,
\end{align}
with the initial condition $\mathcal{G}_{ij}(x=1)=G_{ij}$,
we obtain the two-point functions:
\begin{widetext}
\begin{footnotesize}
\begin{align}
&\mathcal{G}^{-}(x)=|x|^{-2\Delta}
\begin{pmatrix}
 0.0976 |x|^{102.841\alpha} & 0 & 0 & 0 \\
 0 & 0.230 |x|^{-81.288\alpha} & 0 & 0 \\
 0 & 0 & 0 & -5.734 |x|^{13.704\alpha} \\
 0 & 0 & -5.734 |x|^{13.704\alpha} & -(26.004 +255.316\alpha  \log |x| )|x|^{13.704\alpha} \\
\end{pmatrix}
+{O}(\delta)\,.
\end{align}
\end{footnotesize}
\end{widetext}
The last diagonal entry exactly recovers the result for $\tilde{\mathcal{G}}_{00}$ in Eq.\,\eqref{eq:G00}.

\section{Technicality of geometric phase}
\label{app:geophase}

For non-Hermitian systems defined by $\mathbb{D}$, we introduce the (unnormalized) right and left eigenvectors, $\{v^a\}$ and $\{u^a\}$ respectively, defined by
\begin{align}
\label{eq:defineUV2}
& \mathbb{D}  \, v^a=E_a \,v^a\,,\quad
(u^a)^{\mathrm{T}}  \mathbb{D}  =E_a \,(u^a)^{\mathrm{T}}\,,
\end{align}
Note that the left eigenvector is different from the definition in (\ref{eq:defineUV}).
Here $\mathbb{D}$ is the original dilatation matrix, e.g.,
 Eq.\,(\ref{eq:oldDila1}) for the operator sector in Eq.\,(\ref{eq:newbaselen4}). 
Generally, $\{u^a\}$ and $\{v^a\}$ are distinct for non-symmetric $\mathbb{D}$. 
These two sets of eigenvectors are  normalized  as follows:
\begin{align}
\label{eq:norm-nonsym}
(\hat{u}^a)^{\mathrm{T}}:=\frac{(u^a)^{\mathrm{T}}}{\sqrt{(u^a)^{\mathrm{T}} v^a}   }\,,\quad
\hat{v}^a:=\frac{v^a}{\sqrt{(u^a)^{\mathrm{T}} v^a} }\,.
\end{align}
This normalization ensures the biorthonormality condition:
\begin{align}
\label{eq:uv-Delta}
(\hat{u}^a)^{\mathrm{T}}\hat{v}^b=\delta^{ab}\,.
\end{align}
Actually the unnormalized eigenvectors already satisfy 
the orthogonality condition, since for a non-degenerate system with $E_a\neq E_b$:
\begin{align}
(u^a)^{\mathrm{T}}(E_a-E_b)v^b
=(u^a)^{\mathrm{T}} \mathbb{D} v^b-(u^a)^{\mathrm{T}} \mathbb{D} v^b=0\,.
\end{align}
For the dilatation matrix $\mathbb{D}$ that are regular in $N_c$ (which is true for the original dilatation matrix but not for the symmetric $\mathbb{H}$),
the denominators in Eq.\,(\ref{eq:norm-nonsym}) are non-zero except at the EP.
In other words, this normalization guarantees that EPs are 
the only singular points of the eigenvectors. 
Therefore, any non-trivial geometric phase originates exclusively from encircling EPs.

Furthermore, this normalization guarantees that the 
monodromy matrices derived from the left and right eigenvectors are closely related.
To be concrete, write the two monodromy matrices defined by Eq.\,(\ref{eq:defineM}) 
for $v^a$ and $u^a$ respectively:
\begin{align}
\hat{v}^a\big|_{\mathrm{fin}}=\mathbb{M}_{ab}(\mathcal{C})\cdot \hat{v}^b\big|_{\mathrm{ini}}\,,
\quad
(\hat{u}^a)^{\mathrm{T}}\big|_{\mathrm{fin}}=
(\hat{u}^b)^{\mathrm{T}}\big|_{\mathrm{ini}} \cdot \mathbb{N}_{ba}(\mathcal{C})\,.
\nonumber
\end{align}
The biorthonormality condition Eq.\,(\ref{eq:uv-Delta}) dictates that $\mathbb{N} \cdot\mathbb{M}=\mathbbm{1}$.
Also, recall that the non-Abelian Berry connection responsible for the parallel transport
of $v^a$ is given by Eq.\,(\ref{eq:omega2}) \cite{vanderbilt2018berry}.
One can similarly consider the connection responsible for the parallel transport
of $(u^a)^{\mathrm{T}}$ by interchanging $u\leftrightarrow v$ in Eq.\,(\ref{eq:omega2}).
From Eq.\,(\ref{eq:uv-Delta}), we know that they only differ by a minus sign.

One can define the single-encircling paths for the nine other EPs  
analogous to the path shown in Figure~\ref{fig:contour}, 
and accordingly obtain nine other monodromy matrices.
All these monodromy matrices satisfy the property that
their quartic powers  equal the identity matrix.
One can verify that deformed contours yield the same monodromy result as long as
they are topologically equivalent, where the `equivalent' refers to sharing the same
base point, encircling the same EP, and having the same winding number. 
Therefore, the monodromy matrices are (non-abelian) geometric phases induced by
the variation of the parameter $N_c$.

For a contour obtained by gluing two single encircling paths $\mathcal{C}(a)$
and $\mathcal{C}(b)$ together, the associated monodromy is just the product
$\mathbb{M}(a)\cdot\mathbb{M}(b)$. The group of monodromy matrices for arbitrary
closed paths can be generated by the 10 matrices associated with the single-encircling paths.
For the system in Eq.\,(\ref{eq:oldDila1}), this group realizes the permutation between any two
eigenvectors out of the four.
Translated back to the YM theory, this results in a complicated interchange pattern between
correlation functions.

\bibliography{ff}

\end{document}